\documentclass[aps,prd, preprint, superscriptaddress,preprintnumbers, amsmath,amssymb,amsfonts, nofootinbib,longbibliography]{revtex4-1}

\usepackage{graphicx}
\graphicspath{{./figs/}}
\usepackage[dvipsnames]{xcolor}
\usepackage{hyperref}
\usepackage{xspace}
\usepackage{ifdraft}
\usepackage{amsmath,mathdots} 
\usepackage{epstopdf}
\usepackage{slashed}
\usepackage{mathtools}
\usepackage[framemethod=default]{mdframed}
\newmdenv[skipabove=7pt,
skipbelow=7pt,
rightline=false,
leftline=false,
topline=false,
bottomline=false,
backgroundcolor=gray!15,
linecolor=gray,
innerleftmargin=5pt,
innerrightmargin=5pt,
innertopmargin=5pt,
innerbottommargin=5pt,
leftmargin=0cm,
rightmargin=0cm,
linewidth=4pt]{eBox}

\usepackage{tikz-feynman}

\tikzfeynmanset{compat=1.1.0}
\tikzfeynmanset{/tikzfeynman/warn luatex = false}

\definecolor{nhpRed}{RGB}{161,0,0}
\definecolor{nhp4}{RGB}{203, 4, 31}
\definecolor{nhp3}{RGB}{244,99,30}
\definecolor{nhp2}{RGB}{255,159,0}
\definecolor{nhp1}{RGB}{48,152,152}
\definecolor{nhpBlue}{RGB}{0,100,144}
\definecolor{nhpBlue5}{RGB}{200,0,0}
\definecolor{nhpBlue1}{RGB}{150,45,50}
\definecolor{nhpBlue2}{RGB}{100,70,100}
\definecolor{nhpBlue3}{RGB}{50,95,150}
\definecolor{nhpBlue4}{RGB}{0,120,200}
\definecolor{jaxoblue}{HTML}{0086FF}
\definecolor{cutred}{RGB}{219,56,49}
\definecolor{hgreen}{RGB}{25,176,146}
\definecolor{hgreen1}{RGB}{175,230,175}
\definecolor{hblue}{RGB}{52,152,219}
\definecolor{hblue1}{RGB}{255,255,166}
\definecolor{hred}{RGB}{216,83,117}
\definecolor{hred1}{RGB}{255,155,155}
\definecolor{cutred}{RGB}{219,56,49}
\definecolor{hgrey4}{RGB}{75,75,75}
\definecolor{hgrey5}{RGB}{50,50,50}
\definecolor{hgrey3}{RGB}{100,100,100}
\definecolor{hgrey}{RGB}{125,125,125}
\definecolor{hgrey2}{RGB}{125,125,125}
\definecolor{hgrey1}{RGB}{150,150,150}
\definecolor{hgrey0}{RGB}{200,200,200}
\definecolor{darkgreen}{RGB}{255,153,153}

\usepackage{xcolor}
\hypersetup{
    colorlinks,
    linkcolor={nhpRed},
    citecolor={nhpBlue},
    urlcolor={nhpBlue}
}

\definecolor{cutred}{RGB}{219,56,49}

\newcommand{\pentaTri}{ {
\scalebox{1}{\begin{tikzpicture}[baseline=-2]
\begin{feynman}
\vertex (mid1) at (-1.2,.6);
\vertex (mid2) at (0,.6);
\vertex (mid3) at (1.2,.6);
\vertex (mid4) at (-1.2,-.6);
\vertex (mid5) at (0,-.6);
\vertex (mid6) at (1.2,-.6);
\vertex (mid7) at (.6,0){};
\vertex (a1) at (-1.5,.9);
\vertex (a2) at (1.5,.9);
\vertex (a3) at (-1.5,-.9);
\vertex (a4) at (1.5,-.9);
\diagram{
(mid1) --[ultra thick](mid2),
(a1) --[ultra thick](mid1),
(a3) --[ultra thick](mid4),
(a2) --[ultra thick](mid3),
(a4) --[ultra thick](mid6),
(mid1) --[ultra thick](mid4),
(mid4) --[ultra thick](mid5),
(mid3) --[ultra thick](mid5),
(mid5) --[ultra thick](mid6),
(mid3) --[ultra thick](mid2),
(mid7) --[ultra thick](mid6),
(mid2) --[ultra thick](mid7),
};
\vertex (mid1a) [dot,opacity=0.5, scale=2,nhpBlue4]  at (-1.2,.6){};
\vertex (mid2a) [dot,opacity=0.5, scale=2,nhpBlue5]  at (0,.6){};
\vertex (mid3a) [dot,opacity=0.5, scale=2,nhpBlue4]  at (1.2,.6){};
\vertex (mid4a) [dot,opacity=0.5, scale=2,nhpBlue4]  at (-1.2,-.6){};
\vertex (mid5a) [dot,opacity=0.5, scale=2,nhpBlue5]  at (0,-.6){};
\vertex (mid6a) [dot,opacity=0.5, scale=2,nhpBlue4]  at (1.2,-.6){};
\end{feynman}
\end{tikzpicture}}
}
}

\newcommand{\nGon}{ {
\scalebox{1}{\begin{tikzpicture}[baseline=-2]
\begin{feynman}
\vertex (mid1) at (.7,0);
\vertex (mid2) at (.35,.606);
\vertex (mid3) at (-.35,.606);
\vertex (mid4) at (-.7,0);
\vertex (mid5) at (-.35,-.606);
\vertex (mid6) at (.35,-.606);
\vertex (a1) at (1.7,0){$n\!-\!1$};
\vertex (a2) at (.7,1.212){$n$};
\vertex (a3) at (-.7,1.212){$1$};
\vertex (a4) at (-1.4,0){$2$};
\diagram{
(mid1) -- [ultra thick](a1),
(mid2) -- [ultra thick](a2),
(mid3) -- [ultra thick](a3),
(mid4) -- [ultra thick](a4),
(mid1) --[ultra thick](mid2),
(mid2) --[ultra thick](mid3),
(mid3) --[ultra thick](mid4),
(mid4) --[ultra thick](mid5),
(mid5) --[scalar,ultra thick](mid6),
(mid6) --[ultra thick](mid1),
}; 
\vertex (mid1) [dot,opacity=0.5, scale=2,nhpBlue4] at (.7,0){};
\vertex (mid2)[dot,opacity=0.5, scale=2,nhpBlue4]  at (.35,.606){};
\vertex (mid3) [dot,opacity=0.5, scale=2,nhpBlue4] at (-.35,.606){};
\vertex (mid4) [dot,opacity=0.5, scale=2,nhpBlue4] at (-.7,0){};
\vertex (mid5) [dot,opacity=0.5, scale=2,nhpBlue4] at (-.35,-.606){};
\vertex (mid6) [dot,opacity=0.5, scale=2,nhpBlue4] at (.35,-.606){};
\end{feynman}
\end{tikzpicture}}
}
}

\newcommand{\halfLadder}[2]{ {
\scalebox{1}{\begin{tikzpicture}[baseline=8]
\begin{feynman}
\vertex (mid1) at (0,0);
\vertex (mid2) at (.7,0);
\vertex (mid3) at (1,0);
\vertex (mid3a) at (1.5,.4){$\cdots$};
\vertex (mid4) at (1.8,0);
\vertex (mid4a) at (2.1,0);
\vertex (a6) at (3,0){$n$};
\vertex (a1) at (-0.9,0){1};
\vertex (a2) at (0,.9){#1};
\vertex (a3) at (.7,.9){#2};
\vertex (a4) at (2.1,.9){$n-1$};
\diagram{
(mid1) -- [ultra thick](a1),
(mid1) -- [ultra thick](a2),
(mid2) -- [ultra thick](a3),
(mid4a) -- [ultra thick](a4),
(mid4a) -- [ultra thick](mid4),
(mid4a) -- [ultra thick](a6),
(mid1) --[ultra thick](mid2),
(mid2) --[ultra thick](mid3),
(mid3) --[scalar,ultra thick](mid4),
};
\vertex (mid1)[dot,opacity=0.5, scale=2,nhpBlue4]  at (0,0){};
\vertex (mid2) [dot,opacity=0.5, scale=2,nhpBlue4] at (.7,0){};
\vertex (mid4a)[dot,opacity=0.5, scale=2,nhpBlue4]  at (2.1,0){};
\end{feynman}
\end{tikzpicture}}
}
}

\newcommand{\doubleBox}{ {
\scalebox{1}{\begin{tikzpicture}[baseline=-2]
\begin{feynman}
\vertex (mid1) at (-1.2,.6);
\vertex (mid2) at (0,.6);
\vertex (mid3) at (1.2,.6);
\vertex (mid4) at (-1.2,-.6);
\vertex (mid5) at (0,-.6);
\vertex (mid6) at (1.2,-.6);
\vertex (a1) at (-1.5,.9);
\vertex (a2) at (1.5,.9);
\vertex (a3) at (-1.5,-.9);
\vertex (a4) at (1.5,-.9);
\diagram{
(mid1) --[ultra thick](mid2),
(a1) --[ultra thick](mid1),
(a3) --[ultra thick](mid4),
(a2) --[ultra thick](mid3),
(a4) --[ultra thick](mid6),
(mid1) --[ultra thick](mid4),
(mid4) --[ultra thick](mid5),
(mid2) --[ultra thick](mid5),
(mid5) --[ultra thick](mid6),
(mid3) --[ultra thick](mid2),
(mid3) --[ultra thick](mid6),
};
\vertex (mid1a) [dot,opacity=0.5, scale=2,nhpBlue4]  at (-1.2,.6){};
\vertex (mid2a) [dot,opacity=0.5, scale=2,nhpBlue5]  at (0,.6){};
\vertex (mid3a) [dot,opacity=0.5, scale=2,nhpBlue4]  at (1.2,.6){};
\vertex (mid4a) [dot,opacity=0.5, scale=2,nhpBlue4]  at (-1.2,-.6){};
\vertex (mid5a) [dot,opacity=0.5, scale=2,nhpBlue5]  at (0,-.6){};
\vertex (mid6a) [dot,opacity=0.5, scale=2,nhpBlue4]  at (1.2,-.6){};
\end{feynman}
\end{tikzpicture}}
}
}

\newcommand{\factGraphsTrip}[6]{ {
\scalebox{0.85}{\begin{tikzpicture}[baseline=-3]
\begin{feynman}
\vertex [dot, scale=1.08](mid1) at (-.6,0) {$\alpha^{n+} $};
\vertex [dot, scale=1,white](mid1a) at (-.6,0) {$\alpha^{n+} $};
\vertex [dot, scale=5.3,opacity=0.4,#1](mid1a) at (-.6,0) {$ $};
\vertex (mid1b) at (-.6,0) {#3};
\vertex [dot, scale=1.08](mid2) at (.6,0) {$\alpha^{n+} $};
\vertex [dot, scale=1,white](mid2a) at (.6,0) {$\alpha^{n+} $};
\vertex [dot, scale=5.3,opacity=0.4,#2](mid2a) at (.6,0) {$ $};
\vertex (mid2b) at (.6,0) {#4};
\vertex (a1) at (0,.-.4) ;
\vertex (a2) at (0,.4) ;
\vertex [dot, scale=1.08](mid4) at (1.8,0) {$\alpha^{n+} $};
\vertex [dot, scale=1,white](mid3a) at (1.8,0) {$\alpha^{n+} $};
\vertex [dot, scale=5.3,opacity=0.4,#6](mid3a) at (1.8,0) {$ $};
\vertex (mid3b) at (1.8,0) {#5};
\vertex (a3) at (1.2,.-.4) ;
\vertex (a4) at (1.2,.4) ;
\vertex (mid3) at (0,0) {} ;
\diagram{
(mid1) --[thick](mid2),
(mid4) --[thick](mid2),
(a1) --[thick,scalar,nhpRed](a2),
(a3) --[thick,scalar,nhpRed](a4),
};
\end{feynman}
\end{tikzpicture}}
}
}

\newcommand{\factGraphsVert}[4]{ {
\scalebox{0.85}{\begin{tikzpicture}[baseline=-3]
\begin{feynman}
\vertex [dot, scale=1.08](mid1) at (0,-.6) {$\alpha^{n+} $};
\vertex [dot, scale=1,white](mid1a) at (0,-.6) {$\alpha^{n+} $};
\vertex [dot, scale=5.3,opacity=0.4,#1](mid1a) at (0,-.6) {$ $};
\vertex (mid1b) at (0,-.6) {#3};
\vertex [dot, scale=1.08](mid2) at (0,.6) {$\alpha^{n+} $};
\vertex [dot, scale=1,white](mid2a) at (0,.6) {$\alpha^{n+} $};
\vertex [dot, scale=5.3,opacity=0.4,#2](mid2a) at (0,.6) {$ $};
\vertex (mid2b) at (0,.6) {#4};
\vertex (a1) at (-.4,0) ;
\vertex (a2) at (.4,0) ;
\vertex (mid3) at (0,0) {} ;
\diagram{
(mid1) --[thick](mid2),
(a1) --[thick,scalar,nhpRed](a2),
};
\end{feynman}
\end{tikzpicture}}
}
}

\newcommand{\factGraphsArb}[1]{ {
\scalebox{0.85}{\begin{tikzpicture}[baseline=-2]
\begin{feynman}
\vertex [dot, scale=1.05](mid1) at (-1.5,0) {$D^{2n+}F^4$};
\vertex (b1) at (-2.8,.6) {$\,\,$};
\vertex (b1a) at (-3.1,.6) {$n\!-\!1$};
\vertex (b2) at (-2.8,.1) {$\vdots$};
\vertex (b3) at (-2.8,-.6) {$1$};
\vertex (b4) at (-.2,.6) {$n$};
\vertex (b6) at (-.2,.1) {$\vdots$};
\vertex (b5) at (-.2,-.6) {$\,\,\,$};
\vertex (b7)  at (.2,-.6) {$2n\!-\!2$};
\vertex [dot, scale=1,hgrey0](mid1a) at (-1.5,0) {$D^{2n+}F^4$};
\vertex (mid1b) at (-1.5,0) {#1};
\diagram{
(mid1a) --[ultra thick](b1),
(mid1a) --[ultra thick](b3),
(mid1a) --[ultra thick](b4),
(mid1a) --[ultra thick](b5),
};
\end{feynman}
\end{tikzpicture}}
}
}

\newcommand{\factGraphsArbLeglessG}[1]{ {
\scalebox{1}{\begin{tikzpicture}[baseline=-4]
\begin{feynman}
\vertex [dot, scale=1.05](mid1) at (-1.5,0) {$D^{2n+}F$};
\vertex [dot, scale=1,hgrey0](mid1a) at (-1.5,0) {$D^{2n+}F$};
\vertex (mid1b) at (-1.5,0) {#1};
\vertex (b1) at (-2.6,.8) {$2$};
\vertex (b2) at (-2.8,0) {$1$};
\vertex (b3) at (-2.6,-.8) {$n$};
\vertex (b6) at (-.6,.1) {$\vdots$};
\diagram{
(mid1a) --[ultra thick](b1),
(mid1a) --[ultra thick](b2),
(mid1a) --[ultra thick](b3),
};
\end{feynman}
\end{tikzpicture}}
}
}

\newcommand{\factGraphsArbLegless}[2]{ {
\scalebox{1}{\begin{tikzpicture}[baseline=-4]
\begin{feynman}
\vertex [dot, scale=1.05](mid1) at (-1.5,0) {$D^{2n+}F$};
\vertex [dot, scale=1,gray!10](mid1a) at (-1.5,0) {$D^{2n+}F$};
\vertex [dot, opacity=#2,scale=1,nhpBlue5](mid1a) at (-1.5,0) {$\qquad \quad\,$};
\vertex (mid1b) at (-1.5,0) {#1};
\end{feynman}
\end{tikzpicture}}
}
}

\newcommand{\factGraphsEight}[1]{ {
\scalebox{0.85}{\begin{tikzpicture}[baseline=-2]
\begin{feynman}
\vertex [dot, scale=1.05](mid1) at (-1.5,0) {$D^{2n+}F^4$};
\vertex (b7) at (-2.8,.75) {4};
\vertex (b1) at (-2.8,.25) {3};
\vertex (b2) at (-2.8,-.25) {2};
\vertex (b3) at (-2.8,-.75) {1};
\vertex (b4) at (-.2,.75) {5};
\vertex (b6) at (-.2,.25) {6};
\vertex (b5) at (-.2,-.25) {7};
\vertex (b8) at (-.2,-.75) {8};
\vertex [dot, scale=1,hgrey0](mid1a) at (-1.5,0) {$D^{2n+}F^4$};
\vertex (mid1b) at (-1.5,0) {#1};
\diagram{
(mid1a) --[ultra thick](b1),
(mid1a) --[ultra thick](b2),
(mid1a) --[ultra thick](b3),
(mid1a) --[ultra thick](b4),
(mid1a) --[ultra thick](b6),
(mid1a) --[ultra thick](b5),
(mid1a) --[ultra thick](b7),
(mid1a) --[ultra thick](b8),
};
\end{feynman}
\end{tikzpicture}}
}
}

\newcommand{\factGraphsSix}[1]{ {
\scalebox{0.85}{\begin{tikzpicture}[baseline=-2]
\begin{feynman}
\vertex [dot, scale=1.05](mid1) at (-1.5,0) {$D^{2n+}F^4$};
\vertex (b1) at (-2.8,.6) {3};
\vertex (b2) at (-2.8,0) {2};
\vertex (b3) at (-2.8,-.6) {1};
\vertex (b4) at (-.2,.6) {4};
\vertex (b6) at (-.2,0) {5};
\vertex (b5) at (-.2,-.6) {6};
\vertex [dot, scale=1,hgrey0](mid1a) at (-1.5,0) {$D^{2n+}F^4$};
\vertex (mid1b) at (-1.5,0) {#1};
\diagram{
(mid1a) --[ultra thick](b1),
(mid1a) --[ultra thick](b2),
(mid1a) --[ultra thick](b3),
(mid1a) --[ultra thick](b4),
(mid1a) --[ultra thick](b6),
(mid1a) --[ultra thick](b5),
};
\end{feynman}
\end{tikzpicture}}
}
}

\newcommand{\factGraphsFive}[6]{ {
\scalebox{0.85}{\begin{tikzpicture}[baseline=-2]
\begin{feynman}
\vertex [dot, scale=1.05](mid1) at (-1.5,0) {$D^{2n+}F^4$};
\vertex (b1) at (-2.8,.6) {#4};
\vertex (b2) at (-2.8,0) {#3};
\vertex (b3) at (-2.8,-.6) {#2};
\vertex (b4) at (-.2,.6) {#5};
\vertex (b5) at (-.2,-.6) {#6};
\vertex [dot, scale=1,hgrey0](mid1a) at (-1.5,0) {$D^{2n+}F^4$};
\vertex (mid1b) at (-1.5,0) {#1};
\diagram{
(mid1a) --[ultra thick](b1),
(mid1a) --[ultra thick](b2),
(mid1a) --[ultra thick](b3),
(mid1a) --[ultra thick](b4),
(mid1a) --[ultra thick](b5),
};
\end{feynman}
\end{tikzpicture}}
}
}

\newcommand{\factGraphsFFFLeglessG}[3]{ {
\scalebox{1}{\begin{tikzpicture}[baseline=-4]
\begin{feynman}
\vertex [dot, scale=1.05](mid1) at (-1.4,0) {$D^{2n}\,F$};
\vertex [dot, scale=1,hgrey0](mid1a) at (-1.4,0) {$D^{2n}\,F$};
\vertex (mid1b) at (-1.4,0) {#1};
\vertex [dot, scale=1.08](mid2) at (.3,0) {$\alpha^{n+} $};
\vertex (a1) at (-.45,-.4) ;
\vertex (a2) at (-.45,.44) ;
\vertex (mid3) at (0,0) {} ;
\diagram{
(mid1) --[ thick](mid2),
(a1) --[thick,scalar,nhpRed](a2),
};
\vertex [dot, scale=1,white](mid2a) at (.3,0) {$\alpha^{n+} $};
\vertex [dot,opacity=0.4, scale=5.2,#2](mid2c) at (.3,0) {};
\vertex (mid2b) at (.3,0) {#3};
\end{feynman}
\end{tikzpicture}}
}
}

\newcommand{\factGraphsFFFLegless}[4]{ {
\scalebox{1}{\begin{tikzpicture}[baseline=-4]
\begin{feynman}
\vertex [dot, scale=1.05](mid1) at (-1.4,0) {$D^{2n}\,F$};
\vertex [dot, scale=1,gray!10](mid1a) at (-1.4,0) {$D^{2n}\,F$};
\vertex [dot, opacity=#4,scale=1,nhpBlue5](mid1a) at (-1.4,0) {$\qquad\,\,\,\,\,$};
\vertex (mid1b) at (-1.4,0) {#1};
\vertex [dot, scale=1.08](mid2) at (.3,0) {$\alpha^{n+} $};
\vertex (a1) at (-.45,-.4) ;
\vertex (a2) at (-.45,.44) ;
\vertex (mid3) at (0,0) {} ;
\diagram{
(mid1) --[ thick](mid2),
(a1) --[thick,scalar,nhpRed](a2),
};
\vertex [dot, scale=1,white](mid2a) at (.3,0) {$\alpha^{n+} $};
\vertex [dot,opacity=0.4, scale=5.2,#2](mid2c) at (.3,0) {};
\vertex (mid2b) at (.3,0) {#3};
\end{feynman}
\end{tikzpicture}}
}
}

\newcommand{\factGraphsFFFSix}[3]{ {
\scalebox{0.85}{\begin{tikzpicture}[baseline=-2]
\begin{feynman}
\vertex [dot, scale=1.05](mid1) at (-1.4,0) {$D^{2n+}F^4$};
\vertex (b1) at (-2.65,.8) {3};
\vertex (b2) at (-2.65,.25) {4};
\vertex (b3) at (-2.65,-.25) {5};
\vertex (b6) at (-2.65,-.8) {6};
\vertex (b4) at (1.5,.6) {2};
\vertex (b5) at (1.5,-.6) {1};
\vertex [dot, scale=1,hgrey0](mid1a) at (-1.4,0) {$D^{2n+}F^4$};
\vertex (mid1b) at (-1.4,0) {#1};
\vertex [dot, scale=1.08](mid2) at (.6,0) {$\alpha^{n++} $};
\vertex (a1) at (-.3,-.48) ;
\vertex (a2) at (-.3,.52) ;
\vertex (mid3) at (0,0) {} ;
\diagram{
(mid1a) --[ultra thick](b1),
(mid1a) --[ultra thick](b2),
(mid1a) --[ultra thick](b3),
(mid1a) --[ultra thick](b6),
(mid2) --[ultra thick](b4),
(mid2) --[ultra thick](b5),
(mid1) --[ultra thick](mid2),
(a1) --[ultra thick,scalar,nhpRed](a2),
};
\vertex [dot, scale=1,white](mid2a) at (.6,0) {$\alpha^{n++} $};
\vertex [dot,opacity=0.5, scale=6.6,#2](mid2c) at (.6,0) {};
\vertex (mid2b) at (.6,0) {#3};
\end{feynman}
\end{tikzpicture}}
}
}

\newcommand{\factGraphsFFF}[8]{ {
\scalebox{0.85}{\begin{tikzpicture}[baseline=-2]
\begin{feynman}
\vertex [dot, scale=1.05](mid1) at (-1.4,0) {$D^{2n+}F^4$};
\vertex (b1) at (-2.65,.6) {#3};
\vertex (b2) at (-2.65,0) {#2};
\vertex (b3) at (-2.65,-.6) {#1};
\vertex (b4) at (1.5,.6) {#4};
\vertex (b5) at (1.5,-.6) {#5};
\vertex [dot, scale=1,hgrey0](mid1a) at (-1.4,0) {$D^{2n+}F^4$};
\vertex (mid1b) at (-1.4,0) {#6};
\vertex [dot, scale=1.08](mid2) at (.6,0) {$\alpha^{n++} $};
\vertex (a1) at (-.3,-.48) ;
\vertex (a2) at (-.3,.52) ;
\vertex (mid3) at (0,0) {} ;
\diagram{
(mid1a) --[ultra thick](b1),
(mid1a) --[ultra thick](b2),
(mid1a) --[ultra thick](b3),
(mid2) --[ultra thick](b4),
(mid2) --[ultra thick](b5),
(mid1) --[ultra thick](mid2),
(a1) --[ultra thick,scalar,nhpRed](a2),
};
\vertex [dot, scale=1,white](mid2a) at (.6,0) {$\alpha^{n++} $};
\vertex [dot,opacity=0.5, scale=6.6,#7](mid2c) at (.6,0) {};
\vertex (mid2b) at (.6,0) {#8};
\end{feynman}
\end{tikzpicture}}
}
}

\newcommand{\tableOpBlend}{ {
\scalebox{1}{\begin{tikzpicture}
\path [line width = 1,draw=black,->]  (-1.5,.6)--(9.25,.6);
\path [line width = 1,draw=black,->]  (-1.1,1)--(-1.1,-3.8);
\path [line width = 1,draw=nhpBlue,rounded corners]   (-1,-3.5)--(-1,.5)--(8,.5);
\path [line width = 1,fill=hblue,opacity=.2,rounded corners]     (-1,-3.5)--(-1,.5)--(9,.5)--(9,-1)--(4,-3.5);
\path [line width = 1,draw=nhpBlue,rounded corners]   (-1,-3.5)--(-1,.5)--(9,.5)--(9,0)--(2,-3.5);
\path [line width = 1,fill=hblue,opacity=.2,rounded corners]     (-1,-3.5)--(-1,.5)--(9,.5)--(9,0)--(2,-3.5);
\path [line width = 1,draw=nhpBlue,rounded corners]   (-1,-2)--(-1,.5)--(7,.5)--(7,0)--(0,-3.5)--(-1,-3.5)--(-1,-2);
\path [line width = 1,fill=hblue,opacity=.2,rounded corners]    (-1,-2)--(-1,.5)--(7,.5)--(7,0)--(0,-3.5)--(-1,-3.5)--(-1,-2);
\path [line width = 1,draw=nhpBlue,rounded corners]  (-1,-2)--(-1,.5)--(5,.5)--(5,0)--(0,-2.5)--(-1,-2.5)--(-1,-2);
\path [line width = 1,fill=hblue,opacity=.2,rounded corners]   (-1,-2)--(-1,.5)--(5,.5)--(5,0)--(0,-2.5)--(-1,-2.5)--(-1,-2);
\path [line width = 1,draw=nhpBlue,rounded corners]  (-1,-2)--(-1,.5)--(3,.5)--(3,0)--(0,-1.5)--(-1,-1.5)--(-1,-.5);
\path [line width = 1,fill=hblue,opacity=.2,rounded corners]   (-1,-2)--(-1,.5)--(3,.5)--(3,0)--(0,-1.5)--(-1,-1.5)--(-1,-.5);
\path [line width = 1,draw=nhpBlue,rounded corners]  (-1,-.5)--(-1,.5)--(1,.5)--(1,0)--(0,-0.5)--(-1,-0.5)--(-1,-.5);
\path [line width = 1,fill=hblue,opacity=.2,rounded corners]   (-1,-.5)--(-1,.5)--(1,.5)--(1,0)--(0,-0.5)--(-1,-0.5)--(-1,-.5);
\path [line width = 1,fill=nhp4,opacity=.3,rounded corners]   (-1,-3.6)--(-1,.5)--(1,.5)--(1,-3.6);
\path [line width = 1,draw=nhpRed,rounded corners]  (-1,-3.6)--(-1,.5)--(1,.5)--(1,-3.6);
\begin{feynman}
\vertex (a1) at (0,0) {$F^2$};
\vertex (a2) at (0,-1) {$D^2F^2$};
\vertex (a3) at (0,-2) {$D^4F^2$};
\vertex (a4) at (0,-3) {$D^6F^2$};
\vertex (b1) at (2,0) {$F^3$};
\vertex (b2) at (2,-1) {$D^2F^3$};
\vertex (b3) at (2,-2) {$D^4F^3$};
\vertex (b4) at (2,-3) {$D^6F^3$};
\vertex (c1) at (4,0) {$F^4$};
\vertex (c2) at (4,-1) {$D^2F^4$};
\vertex (c3) at (4,-2) {$D^4F^4$};
\vertex (c4) at (4,-3) {$D^6F^4$};
\vertex (d1) at (6,0) {$F^5$};
\vertex (d2) at (6,-1) {$D^2F^5$};
\vertex (d3) at (6,-2) {$D^4F^5$};
\vertex (e1) at (8,0) {$F^6$};
\vertex (e2) at (8,-1) {$D^2F^6$};
\vertex (x2) at (9,1) {$F^n$};
\vertex (x1) at (-1.75,-3.5) {$D^{2k}$};
\end{feynman}
\end{tikzpicture}}
}
}

\newcommand{\Zcross}[4]{ {
\scalebox{0.85}{\begin{tikzpicture}[scale=0.7,baseline=-4]]
\begin{feynman}
\vertex (mid1) [dot, scale=3.4*.7]at (0,0) {$F^2$};
\vertex (mid1) [dot, scale=3.3*.7,white]at (0,0) {$F^2$};
\end{feynman}
\path [line width = 1,fill=hgrey0,opacity=.3,draw=nhpBlue]   (-.707,.707)--(.707,-.707)--(-.707,-.707)--(.707,.707)--(-.707,.707);
\path [line width = 1.3,draw=nhpBlue]   (-.707,.707)--(.707,-.707)--(-.707,-.707)--(.707,.707)--(-.707,.707);
\begin{feynman}
\vertex (mid1) [dot, scale=.8]at(-.707,.707) {};
\vertex (mid1) [dot, scale=.8]at(-.707,-.707) {};
\vertex (mid1) [dot, scale=.8]at(.707,-.707) {};
\vertex (mid1) [dot, scale=.8]at(.707,.707) {};
\vertex (mid1) at(-1,1) {#1};
\vertex (mid1) at(-1,-1) {#3};
\vertex (mid1) at(1,-1) {#2};
\vertex (mid1) at(1,1) {#4};
\end{feynman}
\end{tikzpicture}}
}
}

\newcommand{\Zbox}[4]{ {
\scalebox{0.85}{\begin{tikzpicture}[scale=0.7,baseline=-4]]
\begin{feynman}
\vertex (mid1) [dot, scale=3.4*.7]at (0,0) {$F^2$};
\vertex (mid1) [dot, scale=3.3*.7,white]at (0,0) {$F^2$};
\end{feynman}
\path [line width = 1,fill=hgrey0,opacity=.3,draw=nhpBlue]   (-.707,.707)--(-.707,-.707)--(.707,-.707)--(.707,.707)--(-.707,.707);
\path [line width = 1.3,draw=nhpBlue]   (-.707,.707)--(-.707,-.707)--(.707,-.707)--(.707,.707)--(-.707,.707);
\begin{feynman}
\vertex (mid1) [dot, scale=.8]at(-.707,.707) {};
\vertex (mid1) [dot, scale=.8]at(-.707,-.707) {};
\vertex (mid1) [dot, scale=.8]at(.707,-.707) {};
\vertex (mid1) [dot, scale=.8]at(.707,.707) {};
\vertex (mid1) at(-1,1) {#1};
\vertex (mid1) at(-1,-1) {#2};
\vertex (mid1) at(1,-1) {#3};
\vertex (mid1) at(1,1) {#4};
\end{feynman}
\end{tikzpicture}}
}
}

\newcommand{\Zpent}[5]{ {
\scalebox{0.85}{\begin{tikzpicture}[scale=0.7,baseline=-4]]
\begin{feynman}
\vertex (mid1) [dot, scale=3.4*.7]at (0,0) {$F^2$};
\vertex (mid1) [dot, scale=3.3*.7,white]at (0,0) {$F^2$};
\end{feynman}
\path [line width = 1,fill=hgrey0,opacity=.3,draw=nhpBlue]   (0,1)--(0.952,0.309)--(0.588,-0.809)--(-0.588,-0.809)--(-0.952,0.309)--(0,1);
\path [line width = 1.3,draw=nhpBlue]   (0,1)--(0.952,0.309)--(0.588,-0.809)--(-0.588,-0.809)--(-0.952,0.309)--(0,1);
\begin{feynman}
\vertex (mid1) [dot, scale=.8]at(0,1) {};
\vertex (mid1) [dot, scale=.8]at(0.588,-0.809) {};
\vertex (mid1) [dot, scale=.8]at(0.952,0.309){};
\vertex (mid1) [dot, scale=.8]at(-0.952,0.309){};
\vertex (mid1) [dot, scale=.8]at(-0.588,-0.809) {};
\vertex (mid1) at(0,1.4) {#1};
\vertex (mid1) at(-1.3,.5) {#2};
\vertex (mid1) at(-0.85,-1.1) {#3};
\vertex (mid1) at(0.85,-1.1) {#4};
\vertex (mid1) at(1.3,.5) {#5};
\end{feynman}
\end{tikzpicture}}
}
}

\newcommand{\Zpuff}[5]{ {
\scalebox{0.85}{\begin{tikzpicture}[scale=0.7,baseline=-4]]
\begin{feynman}
\vertex (mid1) [dot, scale=3.4*.7]at (0,0) {$F^2$};
\vertex (mid1) [dot, scale=3.3*.7,white]at (0,0) {$F^2$};
\end{feynman}
\path [line width = 1,fill=hgrey0,opacity=.3,draw=nhpBlue]   (0,1)--(0.952,0.309)--(-0.588,-0.809)--(0.588,-0.809)--(-0.952,0.309)--(0,1);
\path [line width = 1.3,draw=nhpBlue]   (0,1)--(0.952,0.309)--(-0.588,-0.809)--(0.588,-0.809)--(-0.952,0.309)--(0,1);
\begin{feynman}
\vertex (mid1) [dot, scale=.8]at(0,1) {};
\vertex (mid1) [dot, scale=.8]at(0.588,-0.809) {};
\vertex (mid1) [dot, scale=.8]at(0.952,0.309){};
\vertex (mid1) [dot, scale=.8]at(-0.952,0.309){};
\vertex (mid1) [dot, scale=.8]at(-0.588,-0.809) {};
\vertex (mid1) at(0,1.4) {#1};
\vertex (mid1) at(-1.3,.5) {#2};
\vertex (mid1) at(-0.85,-1.1) {#3};
\vertex (mid1) at(0.85,-1.1) {#4};
\vertex (mid1) at(1.3,.5) {#5};
\end{feynman}
\end{tikzpicture}}
}
}

\newcommand{\Zstar}[5]{ {
\scalebox{0.85}{\begin{tikzpicture}[scale=0.7,baseline=-4]]
\begin{feynman}
\vertex (mid1) [dot, scale=3.4*.7]at (0,0) {$F^2$};
\vertex (mid1) [dot, scale=3.3*.7,white]at (0,0) {$F^2$};
\end{feynman}
\path [line width = 1,fill=hgrey0,opacity=.3,draw=nhpBlue]   (0,1)--(-0.588,-0.809)--(0.952,0.309)--(-0.952,0.309)--(0.588,-0.809)--(0,1);
\path [line width = 1,fill=white]   (0,-1*.41)--(0.952*.41,-0.309*.41)--(0.588*.41,0.809*.41)--(-0.588*.41,0.809*.41)--(-0.952*.41,-0.309*.41)--(0,-1*.41);
\path [line width = 1.3,draw=nhpBlue]   (0,1)--(-0.588,-0.809)--(0.952,0.309)--(-0.952,0.309)--(0.588,-0.809)--(0,1);
\begin{feynman}
\vertex (mid1) [dot, scale=.8]at(0,1) {};
\vertex (mid1) [dot, scale=.8]at(0.588,-0.809) {};
\vertex (mid1) [dot, scale=.8]at(0.952,0.309){};
\vertex (mid1) [dot, scale=.8]at(-0.952,0.309){};
\vertex (mid1) [dot, scale=.8]at(-0.588,-0.809) {};
\vertex (mid1) at(0,1.4) {#1};
\vertex (mid1) at(-1.3,.5) {#2};
\vertex (mid1) at(-0.85,-1.1) {#3};
\vertex (mid1) at(0.85,-1.1) {#4};
\vertex (mid1) at(1.3,.5) {#5};
\end{feynman}
\end{tikzpicture}}
}
}

\newcommand{\Zzig}[5]{ {
\scalebox{0.85}{\begin{tikzpicture}[scale=0.7,baseline=-4]]
\begin{feynman}
\vertex (mid1) [dot, scale=3.4*.7]at (0,0) {$F^2$};
\vertex (mid1) [dot, scale=3.3*.7,white]at (0,0) {$F^2$};
\end{feynman}
\path [line width = 1,fill=hgrey0,opacity=.3,draw=nhpBlue]   (0,1)--(0.588,-0.809)--(0.952,0.309)--(-0.952,0.309)--(-0.588,-0.809)--(0,1);
\path [line width = 1.3,draw=nhpBlue]   (0,1)--(0.588,-0.809)--(0.952,0.309)--(-0.952,0.309)--(-0.588,-0.809)--(0,1);
\begin{feynman}
\vertex (mid1) [dot, scale=.8]at(0,1) {};
\vertex (mid1) [dot, scale=.8]at(0.588,-0.809) {};
\vertex (mid1) [dot, scale=.8]at(0.952,0.309){};
\vertex (mid1) [dot, scale=.8]at(-0.952,0.309){};
\vertex (mid1) [dot, scale=.8]at(-0.588,-0.809) {};
\vertex (mid1) at(0,1.4) {#1};
\vertex (mid1) at(-1.3,.5) {#2};
\vertex (mid1) at(-0.85,-1.1) {#3};
\vertex (mid1) at(0.85,-1.1) {#4};
\vertex (mid1) at(1.3,.5) {#5};
\end{feynman}
\end{tikzpicture}}
}
}

\definecolor{NUpurple}{RGB}{078,042,132}

\def\sect#1{section~{\ref{#1}}}
\def\fig#1{Fig.~{\ref{#1}}}

\def\spa#1.#2{\left\langle#1\,#2\right\rangle}
\def\spb#1.#2{\left[#1\,#2\right]}
\def\spash#1.#2{\spa{\smash{#1}}.{\smash{#2}}}
\def\spbsh#1.#2{\spb{\smash{#1}}.{\smash{#2}}}
\def\sand#1.#2.#3{%
\left\langle\smash{#1}{\vphantom1}^{-}\right|{#2}%
\left|\smash{#3}{\vphantom1}^{-}\right\rangle}
\def\sandpp#1.#2.#3{%
\left\langle\smash{#1}{\vphantom1}^{+}\right|{#2}%
\left|\smash{#3}{\vphantom1}^{+}\right\rangle}
\def\sandpm#1.#2.#3{%
\left\langle\smash{#1}{\vphantom1}^{+}\right|{#2}%
\left|\smash{#3}{\vphantom1}^{-}\right\rangle}
\def\sandmp#1.#2.#3{%
\left\langle\smash{#1}{\vphantom1}^{-}\right|{#2}%
\left|\smash{#3}{\vphantom1}^{+}\right\rangle}

\def\nn{\nonumber}

\def\eqn#1{eq.~(\ref{#1})}

\def\be{\begin{equation}}
\def\ee{\end{equation}}
\def\bea{\begin{eqnarray}}
\def\eea{\end{eqnarray}}
\def\ba{\begin{eqnarray}}
\def\ea{\end{eqnarray}}

\newcommand{\sv}[2]{ \langle\![#1]\!\rangle^{#2}_{\textbf{SV}} }

\definecolor{NUpurple}{RGB}{078,042,132}

\begin{document}

\author{John Joseph M. Carrasco}
\affiliation{Department of Physics and Astronomy, Northwestern
  University, Evanston, Illinois 60208, USA}
\author{Nicolas H. Pavao}
\affiliation{Department of Physics and Astronomy, Northwestern
  University, Evanston, Illinois 60208, USA}
  \affiliation{SLAC National Accelerator Laboratory, Stanford University, Stanford, CA 94309, USA}
\title{UV Massive Resonance from IR Double Copy Consistency}  
 \begin{abstract}
From the perspective of effective field theory (EFT), Wilson coefficients of the low energy theory are determined by integrating out modes of the full ultraviolet (UV) theory. The spectrum can be in principle resummed if one has access to all available infrared (IR) coefficients at low energies. In this work we show that there exists a general class of consistent massive resonance double-copy (CMRDC) models where UV massive residues are reconstructed through double-copy consistency conditions between the IR Wilson coefficients of the full EFT expansion. Through a color-dual bootstrap, we find surprisingly that double-copy consistency alone introduces the kinematic factors of CMRDC models that soften high energy behavior by exponentiating color-dual contacts. This bootstrap suggests that our massive resonance paradigm is an inevitable consequence of the duality between color and kinematics, thereby providing a path towards emergent UV structure directly from the IR. We then demonstrate how CMRDC models can capture a spectrum of massive modes compatible with general multiplicity, and use Pad\'{e} extrapolation to solve the inverse problem of identifying massive UV resonance from a small number of IR Wilson coefficients.
\end{abstract}
\maketitle
\hypersetup{linkcolor=black}
\hrule
\tableofcontents
\vspace{2.0em}
\hrule
\hypersetup{linkcolor=nhpRed}
\newpage\section{Introduction}
The duality between color and kinematics \cite{Bern:2008qj,Bern:2010ue} and associated double copy construction \cite{Bern:2010ue} has proven to be an effective tool in the calculation of quantum gravity scattering amplitudes to high orders in perturbation theory~\cite{Bern:2012uf,Bern:2013uka,Bern:2017ucb,Bern:2018jmv}.
 In addition, beyond perturbative calculations, recent literature has demonstrated that color-kinematics duality can also place unexpected constraints on all-order higher derivative operators.  
 This has been shown in both in higher derivative vector theories~\cite{Carrasco:2022lbm,Carrasco:2022sck}, effective scalar theories~\cite{Carrasco:2022sck,Brown:2023srz,Li:2023wdm} as well as  constraints on generalizations of the inverse Kawai-Lewellen-Tye (KLT) momentum kernel \cite{Chen:2022shl,Chen:2023dcx}. This growing body of literature concerning the color-dual constraints on effective field theory suggests that graphical organization of scattering amplitudes can be used to bootstrap UV physics directly from IR data \cite{Carrasco:2022lbm}.  Here we suggest that it is precisely this structure that can inform the organization of massive residues in the context of low energy EFT operators.
 
Before proceeding, we note that there are many other studies of encoding massive states\footnote{ See e.g.~refs.~\cite{Chiodaroli2015rdg,Johnson:2020pny, Momeni:2020vvr,Moynihan:2020ejh,Gonzalez:2021bes,Hang:2021oso}  and references therein for examples.} with double-copy construction that is different than what we propose here. Typically, massive double copies approach the problem via local construction of massive modes, $\mathcal{L}_{\text{kin}}=B(\partial^2 - m^2)B$, where kinetic terms are quadratic in the normal field theoretic sense. These approaches apply the double-copy over local massive propagators of the form:
\be  \label{MassiveDouble}
\mathcal{A} = \sum_g \frac{c_g n_g}{d_g - m_g^2} \quad \longrightarrow \quad \mathcal{M} = \sum_g \frac{\tilde{n}_g n_g}{d_g - m_g^2} 
\ee
While this is a potentially insightful approach to understanding the gravitational equivalence principle through the lens of the double copy, our construction here proposes a different paradigm. 
 
The approach we take is informed by the above-mentioned constraints on higher-derivative operators. Because the duality between color and kinematics is agnostic to the spacetime dimension, as we emphasize below, consistent factorization places constraints beyond the renormalizable sector of any theory.  As demonstrated by $Z$-theory~\cite{MafraBCJAmplString,Broedel:2013tta,Carrasco:2016ldy,Carrasco:2016ygv,Mafra:2016mcc,Huang:2016tag,Azevedo:2018dgo}, the duality between color and kinematics applied to cubic graphs encoding massless dynamics, in combination with consistent factorization, can consistently capture massive resonance and emergent ultraviolet behavior characteristic of stringy dynamics  at least through tree-level. From this perspective, direct approaches like \eqn{MassiveDouble} that capture massive residues in the propagators of massive cubic graphs, could be inadvertently sidestepping a key insight of known double-copy models.

Our approach in this work is grounded on the idea that massive residues in amplitudes of ultraviolet complete double-copy theories, $\mathcal{A}^{\text{UV}}$, including gauge theories, are strikingly described in terms of the massless double-copy. In this paradigm, massive residues are encoded in the Wilson coefficients of color-dual higher derivative operators that can be resummed in the ultraviolet:
\be
\mathcal{A}^{\text{UV}} = \sum_g \frac{c_g N_g}{d_g} \quad \longrightarrow \quad  \mathcal{M}^{\text{UV}} = \sum_g \frac{\tilde{n}_g N_g}{d_g}\,
\ee
where the effective kinematic numerators, $N_g$, contain contributions from all order in higher derivative operators,
\begin{equation}
 N_g = n_g + \sum_k c_k (\alpha')^k n^{(k)}_g\,.
\end{equation}
 whose Wilson coefficients $c_k$ are rigidly tuned to produce the desired resonance profile upon resummation. 
 
Our setup is perhaps is not so surprising in principle. However we also suggest there is a straightforward path to resolve the color-dual UV prediction directly from the color-dual IR amplitude, 
\be
\mathcal{A}^{\text{IR}} =  \sum_g \frac{c_g n_g}{d_g} \quad \longrightarrow \quad\mathcal{A}^{\text{UV}} = \sum_g \frac{c_g N_g}{d_g}
\ee
At four-points we identify a simple and suggestive formula for the UV ordered amplitudes that demonstrates two representations; one which highlights the location of massive poles, and one which exposes the exponential behavior that softens the UV. Both are expressed in terms of the IR amplitude, $A^{\text{IR}}$, stripped out as an overall factor:
\begin{eBox}\begin{align}\label{fourMassive}
A^{\text{UV}}_4 &= A^{\text{IR}}_4  \times \prod_{k=1} \frac{P_k(\sigma_2, \sigma_3)}{(s-\mu_k)(t-\mu_k)(u-\mu_k)}  \\
      &=  A^{\text{IR}}_4 \times \prod_{k=1}^{\infty}\exp\big[c_k\,\Omega_k(\sigma_2,\sigma_3) \big] \label{fourExp}
\end{align}
\end{eBox}
where $s$, $t$, $u$ represent the typical four point Mandelstam invariants, $P_k$ and $\Omega_k$ at four-points are simply functions of Mandelstam permutation invariants, 
\begin{equation}
\sigma_2 = \frac{s^2 + t^2+ u^2}{2} \qquad \sigma_3 = \frac{s^3 + t^3 +u^3}{3}\,.
\end{equation}

We should emphasize that our results do not assert that string theory amplitudes are unique, that color-dual amplitudes must play well with supersymmetry,  or that the monodromy relations of open string amplitudes are in some sense generic.  Rather the type of extended structure represented by exponential higher-derivative towers of \eqn{fourExp} can be a generic feature of color-dual low-energy effective field theory. Such towers, for which string theory amplitudes (superstring and bosonic) are exemplars, originate from a simple bootstrap that starts with a minimal amount of information -- simply the duality between color and kinematics -- and emerge\footnote{We suspect additional principles are at play in uniquely specifying known string theory amplitudes.}  due to factorization consistency.

The organization of the paper is as follows. We start with a brief review in \sect{background} of the notation and formalism that we employ throughout the paper. In \sect{bosonicHint}, we begin our study with examples of massive spectra that appear in double-copy consistent theories encoding massive modes, like the $\text{DF}^2+\text{YM}$ theory, followed by the open and closed string in \sect{stringyStructure}. We then explore to what extent the double-copy consistent effective field-theories can be bootstrapped in \sect{sec:SVBootstrap} -- finding surprisingly rigid constraints locking Wilson coefficients to the UV, forcing the emergence of exponential structure whose massive resonance interpretation softens the UV in the dramatic manner we are familiar with from string theory.  This will lead us to suggest the following generic form for the all-multiplicity structure of ordered-amplitudes in double-copy consistent (DCC) theories:
\begin{eBox}
\be \label{primaryFormula}
A^{\text{UV}}_n =A^{\text{IR}}_n\times \prod_{k}\exp\left[c_k\,\Omega^{\text{DCC}}_k \right]
\ee
\end{eBox}
where $\Omega^{\text{DCC}}_k$ are matrix valued operators that promote the color-dual structure of a BCJ $(n-3)!$ basis of IR amplitudes, $A^{\text{IR}}_n$, to double-copy consistent higher derivative amplitudes, $A^{\text{UV}}_n$. Then in \sect{CMRDC} we show how one can begin to use resurgence methods to tune desired resonance profiles beginning from an all multiplicity color-dual UV promotion. We conclude with a discussion of future work in \sect{conclusions}, including some comments on the implications for constructing color-dual integrands for generic theories. 

\newpage\section{Background} \label{background}

We include this section as a brief review to fix a consistent notation and vocabulary around cubic-graph representations, color-dual numerators, ordered amplitudes, and full amplitudes.  For more thorough reviews and tutorials we refer the interested reader to recent reviews on scattering amplituides~\cite{Travaglini:2022uwo,Mizera:2023tfe}, double-copy~\cite{BCJreview, Bern:2022wqg,Adamo:2022dcm,Bern:2023zkg} as well as perturbative string amplitudes~\cite{Berkovits:2022ivl,Mafra:2022wml}. For massless adjoint double-copy theories -- theories describable as $X \otimes Y$ -- we can write full tree-level amplitudes at any multiplicity $m$ in terms of the set of distinctly labeled cubic-graphs with $m$ external legs, $ \mathcal{G}_{3,m}$,
\begin{equation}
\mathcal{A}^{X\otimes Y}_m = \sum_{g \in \mathcal{G}_{3,m}} \frac{n^{X}_gn^{Y}_g}{d_g}
\end{equation}
where $d_g$ is simply the product of massless internal propagators associated with the labeled graph $g$, and both $n^{X}$ and $n^{Y}$ seperately obey Jacobi relations over every internal edge and antisymmetry around acyclic permutations of internal vertices. For example at four-point, the cubic graphs for the three channels obey color-dual Jacobi-like relationships
\begin{equation}
 n^{X}_s = n^{X}_t + n^{X}_u \qquad  n^{Y}_s = n^{Y}_t + n^{Y}_u\,.
\end{equation}
This means that as far as the numerators contributing to the full amplitude are concerned, the number of graphs is overcomplete -- there is a minimal basis.  Going to a minimal basis in $n^{X}$, is called {\em ordering} with respect to the $X$-copy, or color-ordering if $\mathcal{A}$ is  a gauge theory and the $n^{X}$ are entirely given in terms of adjoint structure constants $f^{abc}$.   There will be some minimal basis of $n^{X}_g$ present. Collecting on those basis elements, the resulting coefficients must be gauge invariant and are called ordered amplitudes.  
One can always pick as a basis the half-ladder, or comb-graphs, with leg 1 and leg $m$ at the far left and far right of the graph -- this is a basis of $(m-2)!$ graphs:
\begin{equation}
\mathcal{A}^{X\otimes Y}_m = \sum_{\sigma \in S_{(n-2)}(2,\ldots,m-1)} n^{X}_{(1|\sigma|m)} A^{Y}(1,\sigma,m)\,.
\end{equation}
Here the ordered amplitudes $A^{Y}(1|\sigma|m)$ are entirely given in terms of cubic graphs that share an ordering with $\sigma$,
\begin{equation}
A^{Y}(1,\sigma,m) = \sum_{g \in \mathcal{G}_{3,\sigma}} \text{sig}(g| \mathcal{G}_{3,m}) \frac{n_g}{d_g}
\end{equation}
where $\text{sig}(\ldots)$ encodes the relative signature between the graph's labeling and its canonical labeling in $\mathcal{G}$. For example, consider the four-point amplitude in a minimal basis in $n^{X}$,
\begin{align}
 \mathcal{A}^{X \otimes Y} &= \frac{n^{X}_s n^{Y}_s}{s} +\frac{ (n^{X}_s - n^{X}_u) n^{Y}_t}{t} + \frac{n^{X}_u n^{Y}_u}{u}\\
   &= n^{X}_s A^{Y}(1234) + n^{X}_u A^{Y}(1324)
 \end{align}
where we write $n^{X}_t$ in terms of basis numerators $n^{X}_s$ and $n^{X}_u$,
and the ordered amplitudes are given by,
\begin{align}
A^{X}(1234)&= \frac{n^{Y}_s}{s} + \frac{n^{Y}_t}{t}\\
A^{X}(1324)&= \frac{n^{Y}_u}{u} - \frac{n^{Y}_t}{t} \,.
\end{align}
The ordered amplitudes themselves are related by virtue of algebraic properties of the $n^{Y}$ and these are typically called the Kleiss-Kujif (KK)~\cite{Kleiss:1988ne} and the Bern-Carrasco-Johansson (BCJ)~\cite{Bern:2008qj} relations which allow a reduction to smaller basis. The fact that the $n^{Y}$ also satisfy Jacboi and anti-symmetry allows an $(n-3)!$ or BCJ basis, which means we can write the full amplitude in terms of a product between ordered amplitudes involving a field-theory KLT \cite{KLT, Bern:1998sv} momentum kernel $S$ to remove doubled-pole structure\,
\begin{align}
 \mathcal{A}^{X \otimes Y} &= A^X \otimes A^Y\nn \\
 &= \!\!\!\!\!\!\!\!\!\!  \sum_{\sigma, \rho \in S_{(n-3)}(2,\ldots,n-2)} \!\!\!\!\!  \!\!\!\!\!  A^{X}(1,\sigma,n-1,n) S[\sigma|\rho]  A^{Y}(1,\rho,n,n-1) 
\label{doubleCopySum}
\end{align}
where $S[\sigma|\rho]_1$ can be defined recursively \cite{Bjerrum-Bohr:2010pnr, Carrasco:2016ldy,Carrasco:2016ygv} as 
\begin{equation}
S[A,j|B,j,C]_i = (k_j\!\cdot \! k_{iB})\,S[A|B,C]_i\,.
\end{equation}
This is named after Kawai, Lewellen, and Tye (KLT) who found~\cite{KLT} that one could express closed string amplitudes as sums over products between Chan-Paton stripped open-string amplitudes. 

As  some of the most dramatic all-order examples of the structure we describe today are in the low energy expansion of string amplitudes, it is worth taking a second to identify some moving parts in string amplitudes and how they may appear to obey different rules than field theory amplitudes. String theory Chan-Paton factors are not in the adjoint, so Chan-Paton ordered open-string amplitudes do not obey field-theory relations.  Rather than thinking of Chan-Paton ordered amplitudes as field-theory building blocks (which they are not) it is far more fruitful, from a low-energy EFT perspective at least, to view Chan-Paton dressed amplitudes as field theory double-copies where one copy caries all order in $\alpha'$ corrections as a putative effective field theory, called $Z$-theory.  Indeed, as we will review in \sect{StringyExamples}, given Chan-Paton dressed Z theory amplitudes one can build the so-called svMap order by order in $\alpha'$, so that can write both the open string and closed string as field theory double-copies as follows:
\begin{align}
 \mathcal{A}^{\text{open}}_n &\equiv \mathcal{A}^{\mathbf{Z}\otimes\text{sYM}}=\mathbf{Z}_n\otimes A^{\text{sYM}}_n \\
 \mathcal{A}^{\text{closed}}_n &\equiv \mathcal{A}^{\text{sYM}\otimes \sv{\text{sYM}}{}}= A^{\text{sYM}}_n \otimes  \textbf{svMap}(A^{\text{sYM}}_n)
\end{align}

In string theory the Chan-Paton ordered amplitudes respect an $\alpha'$ dependent generalization~\cite{Stieberger:2009hq,Monodromy} of the field theory relations known as monodromy relations. The monodromy relations limit to field theory relations in the $\alpha' \to 0$ limit. As such we can see a similar $\alpha'$ dependent string-theory KLT kernel holds for string-theory ordered-amplitudes when contracting over the Chan-Paton-stripped orderings of open-strings and $Z$-theory, defined recursively as,
\be  \label{stringKLT}
S[A,j|B,j,C]^{\alpha'}_i = \text{sin}(\alpha' \pi k_j\!\cdot \! k_{iB}) \,S[A|B,C]^{\alpha'}_i \,.
\ee
This will be used for constructing the svMap from Chan-Paton stripped $Z$ theory amplitudes in \sect{StringyExamples}. As noted above, the composite nature of closed string and gravity amplitudes (at tree-level) has been known since the time of KLT, and is by now long established knowledge. However, it has only recently been understood~\cite{Carrasco:2022lbm,Carrasco:2022sck,Chen:2022shl,Chen:2023dcx,Brown:2023srz,Li:2023wdm} that the graphical organization underlying field theory relations of KK and BCJ can also be used to rigidly fix the \emph{a priori} unconstrained freedom of effective field theories. Below we sketch the origin of these constraints.

\newpage\section{Emergent Massive Modes, A Bosonic Hint}\label{bosonicHint}
First let us review the factorization properties of $\text{DF}^2+\text{YM}$-theory, a known color-dual dimension-six theory~\cite{Johansson:2017srf} that plays an integral part in double-copy construction of bosonic and heterotic string theories~\cite{Azevedo:2018dgo}. As we learned in ref.~\cite{Carrasco:2022lbm}, the amplitudes for this theory can be bootstrapped from the bottom up by starting with 
\be\label{YMF3Lag}
\mathcal{L}^{\text{YM}+\text{F}^3} = -\frac{1}{4}( F^{\mu\nu})^2 + \frac{\alpha'}{3} F^{\mu\nu}F^{\nu \rho} F^{\rho \mu}
\ee
and demanding that the amplitudes are double-copy consistent (i.e. both color-dual and factorizable) through five points. Subject to this constraint, the mass-dimension mixing between operators at $\mathcal{O}(\alpha^n)$ in higher derivative corrections introduces a recursion relation between four-point contacts at orders $\alpha'^{n-1}$ and $\alpha'^{n}$, $A^{(n-1)}_4 \Rightarrow A^{(n)}_4$. Explicitly, if we include the color-ordered four-point contact, $A^{(n-1)}_{(1234)}$, which factorizes with the $F^3$-vertex, then we must include an additional contact at one dimension higher, $A^{(n)}_{(1234)}$, that sums with the Yang-Mills vertex: 
\be\label{F3cuts}
{\lim_{s_{45}\to0} }\,\,\factGraphsFive{\large$A_5^{(n)}$}{1}{2}{3}{4}{5}\!\!=\frac{1}{s_{45}}\left[\factGraphsFFF{1}{2}{3}{4}{5}{\large$A_4^{(n-1)}$}{nhpBlue5}{$F^3$}\!\!+\factGraphsFFF{1}{2}{3}{4}{5}{\large$A_{4}^{(n)}$}{nhpBlue4}{$F^2$}\right]
\ee
The precise definition of $A^{(n)}_{(1234)}$ appearing in the factorization above is the following:
\be\label{YMF3Contacts}
\frac{A^{(n)}_{(1234)}}{u} = (F_1F_2)(F_3 F_4) s^{n-1}_{12}+ \text{cyc}(234) 
\ee
where $(F_iF_j) = \text{tr}[F_i F_j]$ is a trace over linearized vector field strengths, $F_i^{\mu\nu} = k_i^\mu \epsilon_i^\nu - k_i^\nu \epsilon_i^\mu$. After setting residual freedom in the bootstrap to zero, one is led precisely to the four-point amplitudes for the  $\text{DF}^2+\text{YM}$ theory by resumming the bootstrapped geometric series:
\be\label{DF2YMAmp}
A_4^{\text{DF}{}^2+\text{YM}} = A^{\text{YM}}_4 + \alpha'A^{F^3}_4 + \alpha'^{\,2} u\left[ \frac{(F_1F_2)(F_3 F_4) }{(\alpha' \, s -1)s} +\frac{(F_1F_4)(F_2 F_3) }{(\alpha' \, t -1)t}+ \frac{(F_1F_3)(F_2 F_4) }{(\alpha' \, u -1)u}  \right]
\ee 
In terms of color-dual kinematic numerators, the four-point $A_4(s,t)$ ordered amplitude is given by,
\be\label{DF2YMAmpNumDef}
A_4^{\text{DF}{}^2+\text{YM}}(s,t) = \left[\frac{n^{\text{DF}{}^2+\text{YM}}_s}{s} +\frac{n^{\text{DF}{}^2+\text{YM}}_t}{t} \right]
\ee
where
\be\label{DF2YMAmpNum}
n^{\text{DF}{}^2+\text{YM}}_g = n_g^{\text{YM}} +\frac{\alpha' n_g^{F^3}+\cdots + \alpha'^4 n_g^{D^4F^4}}{(1-\alpha' s)(1-\alpha' t)(1-\alpha' u)}
\ee
where $n^{\mathcal{O}_i}_s$ are color-dual vector building blocks defined in \cite{Carrasco:2019yyn}. The full color-dressed amplitude is given
\be
\mathcal{A}^{\text{DF}{}^2+\text{YM}} =\frac{c_s n^{\text{DF}{}^2+\text{YM}}_s}{s} +\frac{c_t n^{\text{DF}{}^2+\text{YM}}_t}{t}+\frac{c_u n^{\text{DF}{}^2+\text{YM}}_u}{u}  
\ee
Concordant with \eqn{fourMassive}, we can clearly rewrite this as,
\be
\label{dfMassiveExposure}
\mathcal{A}^{\text{UV}}_4 = \mathcal{A}^{\text{DF}{}^2+\text{YM}}(s,t) = \mathcal{A}^{\text{IR}} \times \frac{1}{(1-\alpha' s)(1-\alpha' t)(1-\alpha' u)}\,.
\ee
We are promoting here a color-dual massless IR theory, given,
\be
\label{dfIR}
\mathcal{A}^{\text{IR}}_4 = \frac{c_s n^{\text{IR}}_s}{s} +\frac{c_t n^{\text{IR}}_t}{t} +\frac{c_u n^{\text{IR}}_u}{u}\,,
\ee
with IR numerators, 
\be
n^{\text{IR}}_g = n_g^{\text{YM}} {(1-\alpha' s)(1-\alpha' t)(1-\alpha' u)}+\alpha' n_g^{F^3}+\cdots + \alpha'^4 n_g^{D^4F^4}\,.
\ee
Concordant with \eqn{fourExp}, it is not so hard to see also that,
\be
\label{dfExpExposure}
\mathcal{A}^{\text{UV}}_4 =  \mathcal{A}^{\text{IR}} \times \prod_{k=1}^{\infty}{\exp{\left[\frac{ \alpha'{}^k}{k}\left(s^k+t^k+u^k\right)\right]}}\,.
\ee
 
The form of \eqn{DF2YMAmpNum}, \eqn{dfMassiveExposure}, and \eqn{dfExpExposure}  demonstrates a central theme of this paper -- namely, that massive residues in double-copy consistent theories are best understood as all-order higher derivative corrections to color-dual numerators that are resummed over \textit{massless poles}. In the case of $\text{DF}{}^2+\text{YM}$ theory, the massive residues that live in $n^{\text{DF}{}^2+\text{YM}}_s$ are an emergent property of demanding double-copy consistency.  Indeed while the IR amplitude in \eqn{dfIR} is certainly color-dual, the tower of higher-derivative operators encoding for the massive resonance is forced upon us for five-points to be color-dual and correctly factorize.  For the remainder of the paper, we elide the proliferation of $\alpha'$ to simplify the expressions, which of course can be reinstated by dimensional analysis. 

Now let us consider a general EFT expansion of Yang-Mills to understand why one might naturally expect massive resonance to fall out of all order consideration of color-dual structure:
\be \label{YMEFTeq}
\mathcal{L}^{\text{YM-EFT}} =F^2+ \sum_{n=1}^\infty \sum_{k=0}^{n-2} D^{2k}F^{n-k}
\ee
At fixed order $\alpha'^n$, these operators will participate in contact terms that appear at order $\alpha'^{n-2}$ in a higher derivative expansion of the theory. If we were only considering the single-copy gauge theories that are double-copy consistent, we could cap the operator expansion at some fixed $\alpha'^n$ for spacetime dimension $D=2n+4$. This is because the bi-adjoint scalar is critical in $D=6$, and thus the dimension of renormalization is unchanged when double-copying with color factors. 

However, in the interest of constructing generic theories from the double-copy, one will in principle need to consider all-order corrections to Yang-Mills. Indeed, while Yang-Mills is critical in $D=4$, double-copying it with itself yields a gravitational theory, that is critical in $D=2$ by naive power counting. Absent enhanced UV cancellation~\cite{Bern:2012gh, Tourkine:2012ip, Bossard:2013rza, Bern:2013qca,Bern:2007hh, Bern:2008pv, Bossard:2011tq, Tourkine:2012ip, Bern:2014sna, Bossard:2011tq,Herrmann:2018dja,Kallosh:2018wzz,Gunaydin:2018kdz,Edison:2019ovj,Kallosh:2023asd,Kallosh:2023thj} from some hidden symmetry (as is the case in some supersymmetric theories of gravity for which the UV divergence is delayed beyond the naive power counting argument), we should expect an infinite tower of operators to become relevant near the UV cutoff scale of the theory. As such, understanding all order in $\alpha'$ corrections to double-copy consistent theories is necessary for probing UV completions. 
\begin{figure}[t]
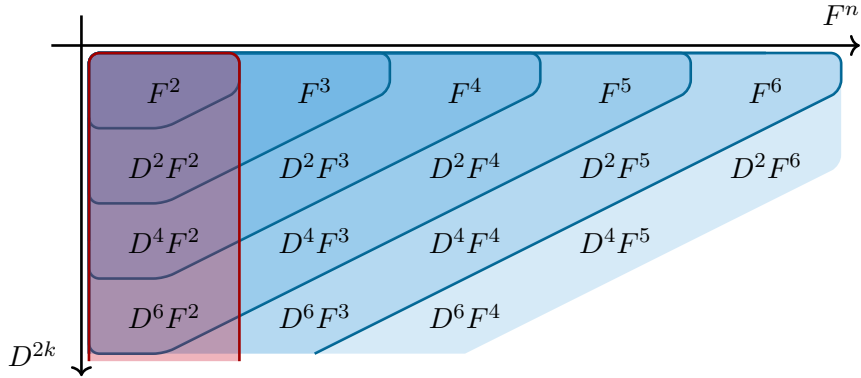

\tableOpBlend
 \caption{We can segment the full Yang-Mills effective expansion by counting derivatives, $D^2$, and field strengths, $F$, in what we call the \textit{color-dual factorization pyramid}. Operators appearing at the lowest rung (lightest blue) in the factorization pyramid are first combined via color-dual constraints on the kinematic numerators (or partial amplitudes) at fixed order $\mathcal{O}(\alpha'^m)$. Then they are woven together with lower orders in $\mathcal{O}(\alpha'^{n<m})$ via factorization. We argue that all order considerations could produce modified kinetic terms, in the same spirit as what was achieved with building $\text{DF}{}^2+\text{YM}$ from double-copy consistent $\text{YM}+\text{F}^3$ amplitudes in \cite{Carrasco:2022lbm}. }\label{colorDualPyramid}
\end{figure}

As we have shown at the beginning of this section, many of these higher-derivative operators exhibit surprising non-trivial relations in their Wilson coefficients. We believe that this could be a hint for a tower of massive resonances needed for all-order double copy consistency. The general setup is something like the sketch depicted in \fig{colorDualPyramid}.  Color-dual constraints mix operators as characterized by their critical dimension at fixed multiplicity, which forces us to consider resummation to higher spin massive resonance (encoded by kinetic corrections in red), and factorization at all multiplicity ties them all together into a consistent theory.  The simplest example is the emergence of a massive mode via demanding color-dual consistency of YM deformed by $F^3$ as we have just discussed.

We call this mechanism for operator mixing in double-copy consistent theories the \textit{color-dual factorization pyramid}. This mechanism, which feeds information from the IR to the UV, can be clearly understood in terms of a color-dual bootstrap. Suppose we started with a four-point vector contact operator, $D^2 F^4\sim \partial^6 A^4$. According to our bootstrap, these are then related to six- and five-field operators through BCJ relations at fixed order in $\alpha'$,
\be \label{OpBundleCK}
\mathcal{O}(\alpha'^3):\quad \partial^4 A^6 \sim \partial^5 A^5 \sim \partial^6 A^4\,,
\ee
By simple power counting, these are then sensitive to six-point factorization constraints of the following form:
\be \label{OpBundleFactorization}
\partial^4 A^6 \sim (\partial^5 A^5) \frac{1}{(\partial A)^2} (\partial A^3) \sim (\partial^6 A^4) \frac{1}{(\partial A)^2} (A^4) \,.
\ee
Upon introducing an infinite tower of such constraints, the operators could in principle resum and encode massive residues in the same fashion as $\text{DF}{}^2+\text{YM}$:
\be \label{nonLocalKinetic}
\mathcal{L}^{\text{YM-EFT}} = F^2+\sum_{n=1}^\infty \sum_{k=0}^{n-2} D^{2k}F^{n-k} \rightarrow F^2+ \prod_{k=1}^\infty (D^2 - \mu_k) F^2+\cdots 
\ee
 The net result would be a $D$-dimensionally color-dual amplitude that captures all the available higher derivative freedom by resumming the masses as follows:
 \begin{eBox}
\be  \label{massiveRes}
A_4^{\text{UV}} = A_4^{\text{IR}} \times \prod_{n=1}^\infty \frac{P_n(\sigma_2,\sigma_3)}{(s-\mu_n)(t-\mu_n)(u-\mu_n)}\,.
\ee
\end{eBox}
 Note that the duality between color and kinematics has been preserved since the four-point UV structure is entirely permutation invariant.  So if the IR ordered amplitudes satisfy KK and BCJ relations the UV theory will as well.  The constraints on Wilson-coefficients appearing in the EFT theory that give rise to this massive spectrum come from factorization consistency of color-dual higher-multiplicity amplitudes down to four-point contacts, as we demonstrate shortly. 
  
  We should point out that this is not a derivation, but an argument for plausibility for why we should expect to see massive resonance structure of the form of \eqn{fourMassive}.  This is in contrast to the sharp derivations we will establish in \sect{sec:SVBootstrap} towards realizing the exponential structure of \eqn{fourExp}.

Before proceeding, let us briefly study the spectrum of the emergent off-shell resonances in \eqn{DF2YMAmp}. Taking residues on the massive poles, and dividing through by the relevant tensor structures, we discover that the fully resummed theory introduces a non-minimally coupled off-shell scalar of the same mass as the vector:
\begin{align}\label{DF2YMResS}
\lim_{s\to 1}\frac{A_4^{\text{DF}{}^2+\text{YM}}}{(F_1F_2)(F_3 F_4)}&=\frac{(t+1)}{ (s -1) }
\\
\lim_{t\to 1}\frac{A_4^{\text{DF}{}^2+\text{YM}}}{(F_1F_4)(F_2 F_3)}&=\frac{(s+1)}{ ( t -1) }\label{DF2YMResT}
\\
\lim_{u\to 1} \frac{A_4^{\text{DF}{}^2+\text{YM}}}{(F_1F_3)(F_2 F_4)}&=\frac{1}{(u -1)} \label{DF2YMResU}
\end{align}
How do we identify the presence of a scalar? The $u$-channel residue is degree zero in Mandelstam invariants, constraining what could be exchanged via factorization to scalar power counting. 

This emergent massive scalar mode in all three channels, along with the vector poles appearing in the $s$ and $t$ channels, are indeed encoded by the $\text{DF}{}^2+\text{YM}$ Lagrangian density, a color-dual dimension six theory first described by Johansson and Nohle \cite{Johansson:2017srf}:
\be\label{DF2YMLag}
\mathcal{L}^{\text{DF}{}^2+\text{YM}} = \frac{1}{2}(D^\mu F^{\mu\nu})^2-\frac{1}{2}m^2 (F^{\mu\nu})^2+\frac{1}{2}(D_\mu \varphi)^2-\frac{1}{2}m^2\varphi^2+\frac{1}{3} F^{\mu\nu}F^{\nu \rho} F^{\rho \mu}+\cdots\,.
\ee

Why is the simple addition of this massive mode compatible with double-copy consistency?  As we will show, the answer is because it enforces the requisite exponential structure on the Wilson-coefficients -- a condition we will establish over the next two sections.

\newpage\section{Consistent Massive Structure via the Single Value Map}\label{stringyStructure}
\label{StringyExamples}
Similar massive spectra to those appearing in $\text{DF}{}^2+\text{YM}$ theory can be also found in the double-copy construction of open and closed string amplitudes. Just as we found in the case of $\text{DF}{}^2+\text{YM}$, massive residues of string amplitudes can encoded as higher derivative corrections in \textit{massless} color-dual numerators. In this section, we explicate this structure and use the low energy EFT expansion of string amplitudes as an exemplar for massive resonance in the double-copy construction. 

We have long known \cite{KLT} that closed string amplitudes can be understood in terms of Chan-Paton ordered (stripped) open string amplitudes
\be\label{CSdouble}
\mathcal{M}^{\text{CS}} =A^{\text{OS}} \overset{~\alpha'}{\otimes}{} A^{\text{OS}}
\ee
where $\overset{~\alpha'}{\otimes}$ is the string momentum kernel of Kawai, Lewellen and Tye (KLT) that works between Chan-Paton stripped quantities which obey the monodromy relations of ordered open strings.  We will refer to such constructions as string double-copies. In addition to the above string double-copy construction relating open and closed strings, the open string amplitudes themselves admit a field theory double copy description.   The open string can be written as
\be\label{OSdouble}
A^{\text{OS}} = Z \otimes A^{\text{YM}} \,
\ee
where $Z$ refers to a putative effective bi-colored scalar theory that carries all order in $\alpha'$ corrections, and $\otimes$ is the field theory double-copy over massless graphs.  As can be seen here, all the massive modes of the superstring are carried by $Z$ theory. 

The $Z$ amplitudes have special properties with respect to one of the color factors.  One set of factors associated with string theory relations, the Chan-Paton factors, involve antisymmetric $f^{ABC}=\text{tr}[t^A,[t^B,t^C]]$ and  symmetric $d^{ABC}=\text{tr}[t^A,\{t^B,t^C\}]$ color-weights.  Ordering along the Chan-Paton factors results in functions that obey string monodromy relations.  The other set of color-factors are simply adjoint structure constants.  Ordering along the adjoint color-factors, which we will call field-theory ordering, results in functions that obey field-theory amplitude relations including the famous $(n-2)!$-basis Kleiss-Kujif associated with graph-by-graph numerator weights that obey anti-symmetry and the $(n-3)!$-basis Bern-Carrasco-Johansson (BCJ) relations associated with numerator weights that also obey adjoint-type Jacobi-like relations.  

The field theory double-copy $\otimes$ can be equivalently expressed as replacing the Yang-Mills color-weights with $Z$ theory color-dual kinematic numerators cubic graph by graph in the full amplitude, or replacing $Z$ theory adjoint-color-weights with color-dual Yang-Mills numerators cubic graph by graph in the full amplitude, or as a relationship between ordered amplitudes contracted with a field-theory KLT momentum kernel.   As we will explain, the $Z$-theory double-copy construction is consistent with our paradigm of massive resonance and the double-copy. We will demonstrate this explicitly by computing some of the low multiplicity matrix elements at four- and five-point below. 

Furthermore, through a little matrix algebra, it is not hard to see that the closed string itself also exhibits via only single valued multiple zeta values~\cite{Schlotterer2012ny,Stieberger:2013wea,Stieberger:2014hba,Schlotterer:2018zce,Vanhove:2018elu,Brown:2019wna} a field theory double copy construction:
\be \label{CSsvPromotion}
\mathcal{M}^{\text{CS}} =\text{sv}(A^{\text{YM}})\otimes A^{\text{YM}} = A^{\text{YM}} \otimes \sv{A^{\text{YM}}}{}
\ee
where we annotate with $\sv{\ldots}{}$ the single-valued (SV) promotion described in~\cite{Carrasco:2022lbm},
\be \label{DefsvPromotion} 
 \sv{X}{} \equiv (Z \overset{~\alpha'}{\otimes} Z) \otimes X \,.
\ee
Note $Z \overset{~\alpha'}{\otimes} Z$ is a doubly-ordered object, where both orderings obey field-theory relations. This promotion lifts color-dual field-theory gauge amplitudes to UV complete field theory gauge amplitudes with an infinite tower of derivatives that encode massive spectra in a manner consistent with color-kinematics and factorization to all multiplicity.  So $\sv{n_g}{}$ obey antisymmetry and Jacobi, defined implicitly so that if
\begin{equation}
\mathcal{X}=\sum_g \frac{c_g {n_g}}{d_g}
\end{equation}
over massless propagators in $d_g$, and $c_g$ is the copy to be ordered along\footnote{Say adjoint-color factors for a gauge-theory.}, then the single-value promotion is given\,
\begin{equation}
\sv{\mathcal{X}}{}=\sum_g \frac{c_g \sv{n_g}{}}{d_g}\,.
\end{equation}
It is worth noting that the SV promotion can be applied interchangeably to both sides of the field theory double-copy. The important takeaway, which we suggest is a general feature of massive resonance in double-copy theories, is that all of the massive modes can be taken as higher derivative corrections on one side of the double-copy.

 \subsection{Z-theory} \label{Ztheory}
The $Z$-theory amplitudes introduced above have an elegant disc integral definition \cite{Broedel:2013tta}:
\be\label{ZtheoryDef}
Z_{(\Pi)}(p_1,p_2,...,p_n)= \int_{D(\Pi)} \frac{dz_{1}dz_{2}\dots dz_{n}}{\text{vol}(\text{SL}(2,\mathbb{R}))} \frac{\prod_{i<j}^n |z_{ij}|^{-s_{ij}}}{z_{p_1p_2}z_{p_2p_3}\dots z_{p_np_1}}
\ee
where $\Pi = (\pi_1,\pi_2,...,\pi_n)$ is the Chan-Paton ordering that obeys monodromy relations and determines the order of integration, and $P=(p_1,p_2,...,p_n)$ is the standard field theory color ordering that functionally specifies the Park-Taylor factor in the denominator of the integation. The full $Z$-theory amplitude is thus recovered by summing over the generators of the $SU(N_1)\times SU(N_2)$ color symmetry:
\be \label{ZtheoryDressing}
\mathbf{\mathcal{Z}}_n = \sum_{\Pi,P}  \text{tr}[t^{\pi_1}t^{\pi_2}\cdots t^{\pi_n}]\text{tr}[t^{p_1}t^{p_2}\cdots t^{p_n}]Z_{\Pi}{(P)} 
\ee
By integrating over the world-sheet, we obtain towers of massive modes consistent with the dual-resonant structure of the open string. 
\paragraph{Four-point} We can perform the integral of \eqn{ZtheoryDef} by fixing three-points in the color ordering, $(z_1,z_{n-1},z_n)\rightarrow (0,1,\infty)$, which allows us to evaluate the disc integral over the remaining $(n-3)$ points. For example, at four-point, the (1234) ordering becomes:
\be \label{Ztheory4ptInt}
Z_{(1234)}{(1234)}=\int_0^1 dz_2\, z_2^{-s_{12}-1}(1-z_2)^{-s_{23}-1}
\ee
To avoid the proliferation of indices, we will introduce a slightly modified version of the the double-ordered graph notation of \cite{Cachazo:2013iea}, where the external points are Chan-Paton ordering, and the internal points are the field theory color ordering. For example, we have:
\be  \label{Ztheory4ptGraphs}
\Zbox{1}{2}{3}{4} = Z_{(1234)}{(\textcolor{nhpBlue}{1234})}\qquad \Zcross{1}{3}{2}{4}=Z_{(1234)}{(\textcolor{nhpBlue}{1324})}
\ee
With this notation in hand, we find that crossing symmetric amplitude of \eqn{Ztheory4ptInt} is the Veneziano amplitude, and the other differs by a factor of $s_{12}/s_{13}$ from BCJ relations:
\be  \label{Ztheory4ptAmps}
\Zbox{1}{2}{3}{4}= \frac{\Gamma(-s_{12})\Gamma(-s_{23})}{\Gamma(-s_{12}-s_{23})}
\qquad
\Zcross{1}{3}{2}{4}= -\frac{\Gamma(1-s_{12})\Gamma(-s_{23})}{\Gamma(1-s_{12}-s_{23})}
\ee
These $Z$-amplitudes have poles located at integer values $s_{ij} = n$, and traditionally can be expressed in their dual resonant form as a sum over poles:
\be\label{Ztheory4ptDualRes}
\Zbox{1}{2}{3}{4} = \sum_{n=0}^\infty \frac{(-1)^{n+1}}{\Gamma(n+1)}\frac{(-s_{23}-n)_n}{s_{12}-n}= \sum_{n=0}^\infty \frac{(-1)^{n+1}}{\Gamma(n+1)}\frac{(-s_{12}-n)_n}{s_{23}-n}
\ee
where the $n$-th massive residue $(x)_n = \Gamma(x+n)/\Gamma(x)$ is simply the Pochhammer symbol. To line them up with massive resonance formula proposed in \eqn{fourMassive} it is best to first compute the SV promotion. Before doing so, we also provide some five-point examples with an eye towards general multiplicity.
\paragraph{Five-point}
Evaluating the five-point $Z$-amplitudes yields a closed-form expression in terms of ${}_3 F_2$ hypergeometric fuctions:
\begin{align} \label{Ztheory5ptA}
\!\!\!\!\Zpent{5}{1}{2}{3}{4}&= {}_3F_2\left[
{-s_{12},\,s_{14},\,-s_{34}\atop
-s_{12}-s_{15},\,-s_{34}-s_{45}}
;1\right]\frac{\Gamma(-s_{15})\Gamma(-s_{12})\Gamma(-s_{34})\Gamma(-s_{45})}{\Gamma(-s_{12}-s_{15})\Gamma(-s_{34}-s_{45})}
\\  \label{Ztheory5ptB}
\!\!\!\!\Zpuff{5}{1}{2}{3}{4}&= -\,{}_3F_2\left[
{1-s_{12},\,s_{14},\,1-s_{34}\atop
1-s_{12}-s_{15},\,1-s_{34}-s_{45}}
;1\right]\frac{\Gamma(-s_{15})\Gamma(1-s_{12})\Gamma(1-s_{34})\Gamma(-s_{45})}{\Gamma(1-s_{12}-s_{15})\Gamma(1-s_{34}-s_{45})}
\end{align}
Using five-point BCJ relations \cite{Bern:2008qj} on the field theory ordering, we obtain two additional functionally distinct $Z$-theory amplitude that appears at five-point:
\be  \label{Ztheory5ptC}
\Zzig{2}{3}{4}{5}{1}=Z_{(12345)}^{(13425)}  = \frac{1}{s_{25}}\left[s_{12 }\,\Zpent{5}{1}{2}{3}{4}+(s_{12}+s_{23})\,\Zpuff{5}{1}{2}{3}{4} \right]
\ee
\be  \label{Ztheory5ptD}
\Zstar{5}{1}{2}{3}{4}=Z_{(12345)}^{(14253)}  = \frac{1}{s_{35}}\left[s_{23}\,\Zpuff{3}{4}{5}{1}{2}+(s_{23}+s_{34})\,\Zzig{2}{3}{4}{5}{1} \right]
\ee
All other $Z$-theory amplitudes are just functional relabelings of these four amplitudes. Above, the intersection points of the field theory ordering indicate the valid region of \textit{massless poles}, but say nothing of restrictions on the massive residues. The star diagram is free of massless poles, but carries an infinite tower of massive residues characteristic of string amplitudes. Evaluating on the $(s_{12},s_{34}) = (m_1,m_2)$ massive poles, we find the following tower of resonances:
\be  \label{Ztheory5ptRes}
\text{Res}\left[\Zpent{5}{1}{2}{3}{4}\right]^{s_{12}=m_1}_{s_{34}=m_2} \!\!\!\!= \frac{(-1)^{m_1+m_2}}{m_1 ! m_2 !}\sum_{n=0}^{\text{min}(m_1,m_2)}\frac{1}{n!}\frac{(-m_1)_n  (s_{14})_n (-m_2)_n}{(-s_{15})_{n-m_1}(-s_{45})_{n-m_2}}
\ee
Note that the series truncates due to the restriction on negative integer values of the Pochhammer symbol, $(-m)_n = 0$ when $n>m$. In appendix \ref{HypergeometricRes}, we have provided helpful background on these functions and their analytic properties. 

\subsection{Single-Value Promotion}
The $Z$-theory amplitudes we have computed in the previous section are sufficient to specify the five-point SV promotion acting on ordered amplitudes.  We need only the string KLT kernel given in \eqn{stringKLT} to proceed.

\paragraph{Four-point}
As there is only one ordered amplitude under the BCJ relations at four-points, the SV promotion is literally multiplication by a permutation invariant $\sigma_V$, for both full amplitudes and individual numerators.  Why $\sigma_V$? It turns out the string KLT product of two complementarily ordered Veneziano factors is the famous Virasoro factor, 
\be  \label{Virasoro}
\sigma_V=\Zbox{1}{2}{3}{4} \text{sin}(\pi s_{14}) \Zcross{1}{3}{3}{4} = \frac{\Gamma(-s_{12})\Gamma(-s_{23})\Gamma(-s_{13})}{\Gamma(s_{12})\Gamma(s_{23})\Gamma(s_{13})} 
\ee
To tease out the infinite product formulae presented in \eqn{primaryFormula}, we will make use of two gamma function identities to expand Virasoro factor of the four-point single-value promotion:
\be \label{GammaIDs}
\Gamma(1+x) = \lim_{n \to \infty }n^x \prod_{k=1}^n \frac{k}{x+k}\quad
\qquad  \Gamma(1+x)= e^{-\gamma x}\prod_{k=2}^\infty \exp\left[\frac{\zeta_k}{k} (-x)^k\right]
\ee
The first expression of \eqn{GammaIDs} allows us to rewrite the Virasoro factor in terms of an infinite product expansion over the massive poles of the theory:
\be  \label{VirasoroMasses}
\sigma_V= \prod_{k=1}^\infty \frac{(s_{12}+k)(s_{23}+k)(s_{13}+k)}{(s_{12}-k)(s_{23}-k)(s_{13}-k)}
\ee
whereas the second allows us to rewrite the Virasoro factor as an infinite product over UV soft exponential factors
\be  \label{VirasoroExps}
\sigma_V= \prod_{k=1}^{\infty}\text{exp}\left[\frac{2\zeta_{2k+1}}{2k+1}(s_{12}^{2k+1}+s_{23}^{2k+1}+s_{13}^{2k+1})\right]
\ee
While this certainly resembles our formula stated in \eqn{primaryFormula}, at this point it could just be a consequence of the simplicity of four-point for which adding higher derivatives to a BCJ basis is just a matter of multiplying by permutation invariants. However, we find surprisingly that the same exponential behavior likewise appears at five-point. 
\paragraph{Five-point}
Performing the same operation as above, but now generalizing to five-point, we obtain the following expression for the SV promotion matrix for orderings $(1P45|1Q45)$:
\be  \label{SVP5point}
\begin{split}
(Z \overset{\alpha'}{\otimes}Z\,  \otimes) = 
&\left[\begin{array}{ccl}\!\!     \Zpent{5}{1}{2}{3}{4} &\!\!      \Zpuff{5}{1}{2}{3}{4} \!    
\\
\!\!      \Zpuff{5}{1}{3}{2}{4} &\!\!     \Zpent{5}{1}{3}{2}{4} \!    
\end{array}\right]\times
\left[\begin{array}{ccl}S[23|23]^{\alpha'}&S[23|32]^{\alpha'}
\\
S[32|23]^{\alpha'}&S[32|32]^{\alpha'}
\end{array}\right]
\\
\times &\left[\begin{array}{ccl}\!\!      \Zpent{4}{1}{2}{3}{5} & \!\!     \Zpuff{4}{1}{2}{3}{5} \!\!     
\\
\!\!      \Zpuff{4}{1}{3}{2}{5} &\!\!     \Zpent{4}{1}{3}{2}{5} \!\!     
\end{array}\,\right]\times
\left[\,\begin{array}{ccl}S[23|23] &\,\,\,\,\,S[23|32]
\\
S[32|23] &\,\,\,\,\,S[32|32]
\end{array}\,\,\,\right]
\end{split}
\ee
While the fully resummed SV promotion is a complicated function of nested integrals over hypergeometric ${}_3F_2$ functions, we can series expand around small $\alpha'$ in the field theory limit, as was done in \cite{Schlotterer2012ny, Broedel:2013tta} using a series of polylogarithmic identities.
 Massaging the results of \cite{Broedel:2013tta} into a form compatible with our definition of the SV promotion, one can show that it exponentiates a countable set of higher derivative matrices:
\be  \label{SVP5pointExponential}
(Z \overset{\alpha'}{\otimes}Z\,  \otimes)= \text{exp}\left[\sum_{n=1}\sum_{k_1,k_2,...,k_n} c_{k_1,k_2,...,k_n} [\Omega_{k_n},[\Omega_{k_{n-1}},\,\cdots ,[\Omega_{k_2},\Omega_{k_1}]]\cdots] \right]
\ee
Above, $\Omega_{2m+3}$ are indexed by odd integers, and take the following form:
\begin{eBox}
\be  \label{HDMatrxiDef}
\Omega_{2m+3} = \left[\begin{array}{ccl}\sigma^{(1|23|45)}_{2m+3}+s_{12}s_{34} X^{2m+3}_{(23|23)} & s_{13}s_{24} X^{2m+3}_{(23|32)} 
\\
 s_{12}s_{34} X^{2m+3}_{(32|23)} &\sigma^{(1|23|45)}_{2m+3}+s_{13}s_{24} X^{2m+3}_{(32|32)}
\end{array}\right]
\ee
\end{eBox}
where we have defined the kinematic factor,
\be \label{PermFactorDef}
\sigma^{(1|23|45)}_{2m+3} = (s_{15}s_{12}s_{25})(s_{15}^2+s_{12}^2+s_{25}^2)^m +(s_{34}s_{45}s_{35})(s_{34}^2+s_{45}^2+s_{35}^2)^m 
\ee
and matched the $X^{2m+3}_{(23|32)}$ and $X^{2m+3}_{(23|23)}$ up to $\mathcal{O}(\alpha'^7)$, which are included the ancillary file. 

For higher order expressions, we refer the reader to \cite{openStringData}. It is maybe not too surprising that the single zeta values exponentiate, as they must consistently factor down to the four-point Virasoro factor, which permits an exponential product expansion as we have shown above. However, it seems at first glance like a small miracle that the exponential form persists to the (single-valued)  multiple zeta value sector which indexes higher-point contacts in the low energy expansion. As we will argue in the next section, we believe that exponentiation is actually a very general property of double-copy consistent amplitudes -- and furthermore expect this to be an all multiplicity feature, not a glitch of the simplicity of beta functions and hypergeometric structure at four- and five-points.

\newpage\section{Double Copy Consistency = Exponentiation}  \label{sec:SVBootstrap}
It is easy to imagine that double-copy consistency admits more freedom in Wilson coefficients than simply the zeta values present in the disc integrals of $Z$-theory and the single-value promotion. Indeed, recent literature has demonstrated that open string amplitudes permit a vast landscape of generalizations consistent with unitarity \cite{Cheung:2023uwn}. As such, here we explore what constraints are placed on Wilson coefficients by the requirement of double-copy consistency, i.e. the duality between color and kinematics and consistent factorization.  We begin with some general four-point color-dual amplitude with a tower of permutation invariants each indexed by a free Wilson coefficient.  We will see how considerations at five and six-points relate these coefficients together.    We will find  that the type of exponentiation exemplified in the single-value map (and indeed in the single massive resonance required to make $\text{YM}+\text{F}^3$ double-copy consistent), as per \eqn{fourExp}, appears universal to double-copy consistent theories in the realm we explore -- those related to the easiest to analyze higher-derivative operators at four-points, those encoded in $\sigma_3^n$ contributions.

\subsection{Five-point}
Guided by the structure of the single-value promotion, we will consider to what extent a color-dual amplitude at four-point acquires further constraints from double copy consistency at five- and six-point. Starting with the four-point EFT expansion for a color-dual vector amplitude, $A^{(0,0)}_4$,
\be   \label{4ptDCCAmp}
A^{(m,n)}_4 =A^{(0,0)}_4 +\sum_{m,n}c_{(m,n)}A^{(m,n)}_4 \qquad A^{(m,n)}_4 \equiv A^{(0,0)} \sigma_3^m \sigma_2^n
\ee
where to be consistent with field theoretic locality we assume $A^{(0,n)}_4=0$ for $n \neq 0$. In what follows, without loss of generality we will fix the ordering of the legs, $(1,n-1, n)$, to mirror our exposition of string amplitudes in the previous section. We are interested in bootstrapping a higher derivative transformation matrix $\Omega^{(m,n)}$ that builds a higher-multiplicity vector amplitude, $A^{(m,n)}$, at mass dimension $3m+2n$ above $A^{(0,0)}$, which consistently factorizes to $A_3 \times A^{(m,n)}_4$:
\be  \label{NptDCCAmpAnsatz}
A^{(m,n)}_{(P)} = \sum_{Q\in S_{(n-3)}}c_{(m,n)}\Omega^{(m,n)}_{(P|Q)} A^{(0)}_{(Q)}\,.
\ee
Furthermore, we want $A^{(m,n)}_{(P)}$ to be double-copy consistent and obey all the same field theory relations as $A^{(0,0)}_{(P)}$. 

Let us emphasize we are talking about the factorizing amplitudes.  There can always be entirely color-dual consistent contact terms that have no factorization channels.  For example, it is sufficient to consider a commutator of $\Omega$'s which when non-vanishing, by definition, will represent contact terms.  Indeed in string theory amplitudes these show up with Wilson coefficients that are multiple zeta values that consistently encode the overlap of the  massive modes responsible for single zeta vlaued Wilson coefficients at four-points. 

Since the matrix elements in each row are related by functional relabelling, a completely general ansatz for factors of $(k_ik_j)^3$ has 70 free parameters. Imposing BCJ relations and factorization constraints on the $s_{12}, s_{34}$ and non-planar $s_{13},s_{24}$ channels, we obtain the following restricted form:
\be  \label{preCutBootstrap5ptMatrix}
\Omega^{(1,0)}_{} = 
\left[\begin{array}{ccl}\sigma_{512}+\sigma_{345}+s_{12}s_{34}X^{(1,0)}_{(23|23)} \quad & s_{13}s_{24}X^{(1,0)}_{(23|32)} 
\\
s_{12}s_{34}X^{(1,0)}_{(32|23)} \quad &\sigma_{513}+\sigma_{245}+s_{13}s_{24}X^{(1,0)}_{(32|32)} \end{array}\right]
\ee
where $X^{(1,0)}_{(23|23)}$ and $X^{(1,0)}_{(23|32)}$ are functional expressions of mandelstams $s_{ij}$ with 5 free parameters each. Finally, imposing the $s_{45}$ and $s_{15}$ cuts, which mixes contributions from the (12345) and (13245) orderings, 
\begin{align}  \label{5ptMatrixCutA}
{\lim_{s_{45}\to0} }\,\,\factGraphsFive{\large$A_5^{(3)}$}{1}{2}{3}{4}{5}\!\!&=c_{(1,0)}\frac{\sigma_{123}}{s_{45}}\,\factGraphsFFF{1}{2}{3}{4}{5}{\large$ A_4$}{nhpBlue4}{\large$A_3$}
\\   \label{5ptMatrixCutB}
{\lim_{s_{15}\to0} }\,\,\factGraphsFive{\large$A_5^{(3)}$}{2}{3}{4}{5}{1}\!\!&=c_{(1,0)}\frac{\sigma_{234}}{s_{15}}\,\factGraphsFFF{2}{3}{4}{5}{1}{\large$ A_4$}{nhpBlue4}{\large$A_3$}
\end{align}
we find surprisingly that there is no remaining freedom in the ansatz -- both $X^{(1,0)}_{(23|23)}$ and $X^{(1,0)}_{(23|32)}$ are completely determined:
\begin{eBox}
\be  \label{5ptMatrix(1,0)XTerms}
X^{(1,0)}_{(23|23)} = -(s_{13}+s_{24}) \qquad X^{(1,0)}_{(23|32)} = (s_{12}+s_{23}+s_{34}+s_{45}+s_{51})
\ee
\end{eBox}
While we did not impose the $s_{23}$-cut, we find that it is consistent with this result. We can perform the same task at one higher order in $\sigma_3$. Starting with an ansatz of degree $(k_ik_j)^6$ with 420 free parameters, and imposing the same factorization constraints on $s_{12}, s_{34}$ and non-planar $s_{13},s_{24}$ channels, we find the transformation matrix takes the following form:
\be \label{preCutBootstrap(2,0)5ptMatrix}
\Omega^{(2,0)}_{} =
\left[\begin{array}{ccl}\sigma^2_{512}+\sigma^2_{345}+s_{12}s_{34}X^{(2,0)}_{(23|23)} \quad & s_{13}s_{24}X^{(2,0)}_{(23|32)} 
\\
s_{12}s_{34}X^{(2,0)}_{(32|23)} \quad &\sigma^2_{513}+\sigma^2_{245}+s_{13}s_{24}X^{(2,0)}_{(32|32)} \end{array}\right]
\ee
where again, $X^{(2,0)}_{(23|23)}$ and $X^{(2,0)}_{(23|32)}$ are functional expressions of mandelstams $s_{ij}$, now each with with 70 free parameters each. Imposing BCJ relations and factorization on the remaining three poles, $s_{45}$ and $s_{15}$, we find again that both $X^{(2,0)}$ are completely determined. Furthermore, they exhibit an iterative structure in terms of the leading order correction matrix:
\be  \label{5ptMatrixRecurrence}
\Omega^{(2,0)}_{(P|Q)}=  (\Omega^{(1,0)}\Omega^{(1,0)})_{(P|Q)} 
\ee
At one order higher in $\sigma_3$, where the starting ansatz has 1430 free parameters, we find that terms proportional to the four-point Wilson coefficient index a transformation matrix with the same iterative structure, $\Omega^{(3,0)} = (\Omega^{(1,0)})^3$. While we anticipate that at higher orders in the EFT expansion there will be five-point contacts that are unconstrained by factorization, we conjecture that after setting that residual freedom to zero, the five-point double-copy consistent amplitudes that factor to $\sigma_3^n A^{(0)}_4 \times A_3$ should obey the following form:
\begin{eBox}
\be  \label{5ptMatrixRecurrenceGuess}
A^{(n,0)}_{(P)} = c_{(n,0)}(\Omega^{(1,0)}\Omega^{(1,0)}\cdots \Omega^{(1,0)})_{(P|Q)} A^{(0)}_{(Q)}
\ee
\end{eBox}
We have checked that this is consistent with the single-zeta values in the single-value promotion, for which $c_{(n,0)} = (2\zeta_3)^n/n!$, in agreement with the exponential behavior of the four-point Virasoro factor appearing in closed string amplitudes in \eqn{VirasoroExps}. Now we will show that double-copy consistency through six-points actually places further constraints on $c_{(n,0)}$, reflective of the exponential structure that appears in the disc integrals of string theory. 
\subsection{Six-point}
We carry out the same procedure as at five point, but now with a $6\times 6$ transformation matrix acting on a BCJ basis $(1|Q|56)$ of six-point amplitudes. At $\mathcal{O}(\alpha'^3)$ above the zeroth order amplitude, we impose all the five-point cuts required by double-copy consistency. One example of a cut that we impose is the $s_{12}$ factorization channel, shown below:
\be   \label{6ptMatrixCut}
{\lim_{s_{12}\to0} }\,\,\factGraphsSix{\large$A_6^{(3)}$}\!\!=\frac{1}{s_{12}}\,\factGraphsFFFSix{\large$A^{(1,0)}_5$}{nhpBlue4}{\large$A_3$}
\ee
After imposing all the two-particle cuts of this form, we then impose BCJ relations on the higher derivative amplitudes. To be double-copy consistent with the five-point amplitudes, the resulting three-particle cuts must then factor as follows:
\be   \label{6ptMatrixSplitCutResultofCK}
{\lim_{s_{123}\to0} }\,\,\factGraphsSix{\large$A_6^{(3)}$}\!\!= c_{(1,0)} \frac{\sigma_{123}+\sigma_{456}}{s_{123}}\left[\factGraphsVert{nhpBlue4}{nhpBlue4}{$\alpha^{0}$}{$\alpha^{0}$}\right]\,,
\ee
where the cartoon stands in for something of order $\alpha^0$ sewn with something of order $\alpha^0$.  It is sufficient to simply be tracking the orders here.
Thus, given the strong evidence for iterative structure of the five-point double-copy consistent amplitudes, we anticipate that in order to be double-copy consistent, the $A_6^{(3n)}$ six-point amplitude must factorize as follows
\begin{eBox}
\be  \label{6ptMatrixSplitCutResultofCKAllorder}
{\lim_{s_{123}\to0} }\,\,\factGraphsSix{\large$A_6^{(3n)}$}\!\!= c_{(n,0)} \frac{(\sigma_{123}+\sigma_{456})^n}{s_{123}}\left[\factGraphsVert{nhpBlue4}{nhpBlue4}{$\alpha^{0}$}{$\alpha^{0}$}\right]
\ee
\end{eBox}
Assuming such a six-point constraint\footnote{A constraint which has been verified through order $\alpha'^{12}$ at six-points via explicit construction.} from double-copy consistency, we can finally impose four-point factorization, from which we obtain a set additional of constraints on the \textit{four-point} Wilson coefficients:
\be \label{6ptMatrixSplitCutConstraint2}
{\lim_{s_{123}\to0} }\,\,\factGraphsSix{\large$A_6^{(6)}$}\!\!=\frac{1}{s_{123}}\left[\factGraphsVert{nhpBlue4}{nhpBlue2}{$\alpha^{0}$}{$\alpha^{6}$}+\factGraphsVert{nhpBlue3}{nhpBlue3}{$\alpha^{3}$}{$\alpha^{3}$}+\factGraphsVert{nhpBlue2}{nhpBlue4}{$\alpha^{6}$}{$\alpha^{0}$}\right]\Rightarrow   c_{(2,0)} = \frac{c_{(1,0)}^2}{2!}
\ee
and likewise at higher order in the derivative expansion:
\begin{align} \label{6ptMatrixSplitCutConstraint3}
{\lim_{s_{123}\to0} }\,\,\factGraphsSix{\large$A_6^{(9)}$}\!\!&=\frac{1}{s_{123}}\left[\factGraphsVert{nhpBlue4}{nhpBlue1}{$\alpha^{0}$}{$\alpha^{9}$}+\factGraphsVert{nhpBlue3}{nhpBlue2}{$\alpha^{3}$}{$\alpha^{6}$}+\factGraphsVert{nhpBlue2}{nhpBlue3}{$\alpha^{6}$}{$\alpha^{3}$}+\factGraphsVert{nhpBlue1}{nhpBlue4}{$\alpha^{9}$}{$\alpha^{0}$}\right]\Rightarrow   c_{(3,0)} = \frac{c_{(1,0)}^3}{3!}
\\  \label{6ptMatrixSplitCutConstraint4}
{\lim_{s_{123}\to0} }\,\,\factGraphsSix{\large$A_6^{(12)}$}\!\!&=\frac{1}{s_{123}}\left[\factGraphsVert{nhpBlue4}{nhpBlue5}{$\alpha^0$}{$\alpha^{12}$}+\factGraphsVert{nhpBlue3}{nhpBlue1}{$\alpha^{3}$}{$\alpha^{9}$}+\factGraphsVert{nhpBlue2}{nhpBlue2}{$\alpha^{6}$}{$\alpha^{6}$}+\factGraphsVert{nhpBlue1}{nhpBlue3}{$\alpha^{9}$}{$\alpha^{3}$}+\factGraphsVert{nhpBlue5}{nhpBlue4}{$\alpha^{12}$}{$\alpha^{0}$}\right]\Rightarrow   c_{(4,0)} = \frac{c_{(1,0)}^4}{4!}
\end{align}
In general, we expect to find the following set of factorization relations for all orders in $\alpha'^n$:
\be  \label{6ptMatrixSplitCutAllOrderConstraint}
\mathcal{O}(\alpha'^{3n}) : \quad \sum_{k=0}^n c_{(n,0)} {n \choose k} \sigma_{(123)}^k  \sigma_{(456)}^{n-k} = \sum_{k=0}^n c_{(k,0)} c_{(n-k,0)} \sigma_{(123)}^k  \sigma_{(456)}^{n-k} 
\ee
Which implies that the Wilson coefficients, $c_{(n,0)}$ at four-point that label matrix elements of the form $\sigma_3^n A_4$, should precisely exponentiate:
\be  \label{Exponetiation4ptC(n,0)}
c_{(n,0)} = \frac{c_{(1,0)}^n}{n!}
\ee
We will now generalize this construction towards the all-multiplicity structure and argue that one should expect double-copy consistency to introduce exponentiation whenever a new contact is added to a color-dual EFT. 
\subsection{All-multiplicity Structure}
We believe that this exponential behavior is likely a universal property of higher-derivative contact terms in double-copy consistent theories, and not just a special property of four-point permutation invariants. In the previous section we just considered how $\sigma_3^n A_4^{\text{YM}}$ type contacts exponentiate. Instead, consider a color-dual $n$-point contact, $X_n$, with dimensionful coupling at $\mathcal{O}(\Lambda)$, that is constructed from transformation matrix $\Omega_n$ acting on an $(n-3)!$ basis of $n$-point vector amplitude at $\mathcal{O}(\Lambda^0)$. Then we can define a general family  of degree $k$ contact operators, $X^{(k)}_n$, as follows,
\be   \label{genColorDualContact}
\factGraphsArbLeglessG{$X_n^{(k)}$}= c_{(k)}\Omega_n^k A_n^{(0)}
\ee
where $c_{(k)}$ is an unconstrained Wilson coefficient fo the $X^k_n$ operator. Now we can promote the product structure of these contacts to higher orders in $\alpha'$ by taking a series of $\text{N}^k \text{Max}$ cuts at successively higher point amplitudes, and imposing color-kinematics at each step:
\begin{equation}  \label{genColorDualRecurrenceBuild}
\mathcal{O}(\Lambda ^{k})\begin{cases}\begin{split}
{\textit{color-dual}\atop \textit{constraints}} \,\,+& \,\,\factGraphsArbLegless{$A_{n+1}^{(k)}$}{.3}\Bigg|^{\text{MC}} \,=\,\factGraphsFFFLeglessG{$\,X^{(k)}_n$}{nhpBlue4}{$A_3$} \quad \Rightarrow \quad A_{n+1}^{(k)} =c_{(k)} \Omega^k_{n+1}A^{(0)}_{n+1}
\\
{\textit{color-dual}\atop \textit{constraints}} \,\,+& \,\,\factGraphsArbLegless{$A_{n+2}^{(k)}$}{.45}\Bigg|^{\text{N}^1\text{MC}} \!\!\! \!=\,\factGraphsFFFLegless{$A_{n+1}^{(k)}$}{nhpBlue4}{$A_3$}{.3}\quad \Rightarrow\quad A_{n+2}^{(k)} =c_{(k)} \Omega^k_{n+2}A^{(0)}_{n+2}
\\
\downarrow \qquad \quad&  \qquad  \qquad\,  \qquad  \qquad \qquad \downarrow  \qquad  \qquad  \qquad  \qquad  \quad \downarrow 
\\
{\textit{color-dual}\atop \textit{constraints}} \,\,+& \,\,\factGraphsArbLegless{$A_{2n-2}^{(k)}$}{.75}\Bigg|^{\text{N}^{n-2}\text{MC}}  \!\! \! \!\! \! \!\! \!\!=\,\factGraphsFFFLegless{$A_{2n-3}^{(k)}$}{nhpBlue4}{$A_3$}{.6}\quad \Rightarrow\quad A_{2n-2}^{(k)} = c_{(k)}\Omega^k_{2n-2}A^{(0)}_{2n-2}
\end{split}\end{cases}
\end{equation}
The blue interaction regions are three-point vertices at $\mathcal{O}(\alpha'^{\,0})$, which do not contribute to the mixing of Wilson coefficients. This is parallel to the first step in checking six-point factorization above, where consistent factorization down to five-point promoted the product structure to the six-point factorization channel. Carrying out double-copy consistency up to $\text{N}^{n-2}\text{MCut}$ level, allows us to now impose a split factorization constraint on the $s_{12...n\!-\!1}$ pole. At this level, the $A^k_{2n-2}$ amplitude would factor into two contact diagrams that set the seed of our recursion:
\begin{align}  \label{genColorDualFactResult}
{\textit{color-dual}\atop \textit{constraints}}\,&\factGraphsArb{\large$A_{2n-2}^{(k)}$} \!\!\!\!\!\!= \frac{1}{s_{12...n\!-\!1}}\sum_{j=0}^k c_{(k)} {k \choose j}\Omega^{\,j}_{(12...n\!-\!1)}\Omega^{k-j}_{(n...2n\!-\!2)}\,\left[\factGraphsVert{nhpBlue4}{nhpBlue4}{\large$A_n$}{\large$A_n$}\right]
\\  \label{genCutConstraint}
{\textit{split-channel}\atop\textit{factorization}}\,&\factGraphsArb{\large$A_{2n-2}^{(k)}$} \!\!\!\!\!\!= \frac{1}{s_{12...n\!-\!1}}\sum_{j=0}^k c_{(j)}c_{(k-j)}\Omega^{\,j}_{(12...n\!-\!1)}\Omega_{(n...2n\!-\!2)}^{k-j}\left[\factGraphsVert{nhpBlue4}{nhpBlue4}{\large$A_n$}{\large$A_n$}\right]
\end{align}
Much like at four-point, all-order color-kinematics and factorization would then require that the Wilson coefficients, $c_{(k)}$, are related to each other by the binomial expansion of $(\Omega_{12... n-1}+\Omega_{n+1... 2n-2})^k$ appearing in \eqn{genColorDualFactResult} -- thus giving following set of relations:
\be  \label{genExponentiation}
c_{(k)} = \frac{1}{k!}c_{(1)}^k
\ee
Since we have made no assumptions about the structure of the color-dual contact beyond it being related to a BCJ basis amplitude, $X_n = \Omega_n A_n$, we state here an all-order ansatz for the all-multiplicity structure of double-copy consistent theories:
\begin{eBox}
\be \label{allMultiplicityFormula}
A^{\text{UV}}_n = \prod_{k}\exp\left[c_k\,\Omega^{\text{DCC}}_k \right]A^{\text{IR}}_n
\ee
\end{eBox}
where $\Omega^{\text{DCC}}_k$ are $(n-3)!\times (n-3)!$ matrices sourcing color-dual contact operators that act on an $(n-3)!$ basis of vector amplitudes at leading order in the EFT expansion. While this remains a conjecture, motivated by structure uncovered in the four-point $\sigma_3$ sector,  directly analyzed at five and six-point, we find it exciting to consider that by making no assumptions about the high energy behavior of an EFT, exponential factors that are compatible with massive modes encoding UV soft behavior could be required at all-multiplicity through double-copy consistency. This is entirely compatible with the position that double-copy consistency as a physical principle is well situated to bootstrap UV  theories directly from low energy effective operators that live in the IR, which we now turn to.

\newpage\section{Consistent Massive Resonance Models}\label{CMRDC}

Effective field theories allow a consistent parameterization of our ignorance beyond data relevant to the scales at hand.  We, especially in the field of scattering amplitudes, are accustomed to considering EFT perturbatively, recognizing a finite bound of validity in relevant energy scales.  In the context of double-copy consistent theories, we have shown how to propagate Wilson-coefficients from four-points to higher multiplicity within the EFT framework.  Both $\text{DF}^2+\text{YM}$ and the SV-map that appears in closed string and heterotic amplitudes provide evidence for a compelling idea: Wilson coefficients that encode massive resonance order by order in $\alpha'$, when constrained by double-copy and factorization, should be compatible with (and in some cases could even be sufficient for) the interpretation of a consistent massive spectrum to propagate from multiplicity to multiplicity solely via Wilson coefficients.  Such a conjecture allows massive-resonance model building in terms of massless kinematics starting at directly from four-points.  

Before we continue with this story let us imagine where such a conjecture could fail. Factorization alone is insufficient to fix contact terms generically.  However, double-copy consistency as we have seen in the emergence of $\text{DF}^2+\text{YM}$ theory establishes uniquely a minimum set of higher-point contact Wilson coefficients in terms of lower-point Wilson coefficients as a bootstrap from Yang-Mills deformed by $F^3$. In that case, factorization and the duality between color and kinematics is sufficient to establish the propagation of massive modes $M$ established as a four-point global prefactor of simply:
\begin{equation}
M^{\text{DF}^2 +\text{YM}}= \frac{1}{(1- s\alpha')(1- t\alpha')(1- u\alpha')}
\end{equation}
against a IR theory that at four-points is color-dual.  

One might ask if factorization and the requirement of color-kinematics duality to all multiplicity may be sufficient to entirely fix the $\alpha'$ expansion of Chan-Paton dressed Z-theory at five-points. This possibility, while attractive, is at least not obvious nor necessary. Starting at $\alpha'^5$ in the five-point amplitude there are local color-dual Wilson-Coefficients required by Z-theory unfixed by factorization to four-points. The presence of multiple zeta values  starting at five-points is a clear sign that any such constraints imposed by all-multiplicity factorization would have to be incredibly non-trivial and seem intimately tied to how massive resonances must interact.  It would be an interesting exercise to learn which Wilson coefficients are ultimately fixed by factorization of color-dual higher-points amplitudes, and what additional constraints are required for consistent interaction of mass-spectra multiplicity by multiplicity.  That of course does not rule out a bootstrap, but suggests that additional principles will be relevant as the multiplicity increases. Given the factorization consistency of color-dual amplitudes with an exponential structure, we are confident a resonance color-dual bootstrap starting from four-points exists, and it will be fascinating to learn the minimal amount of additional principles required to extract Z-theory.

Of course one can imprint massive resonances of some desired  spectrum (from some hierarchically encompassing EFT) into Wilson-coefficients  of color-dual IR four-point amplitudes.  We will discuss how straightforward such a program is, with some words of caution.  We then discuss the much more fascinating knowledge-discovery problem of backing out information about massive resonances, directly from a finite set of Wilson coefficients using resurgence.

\subsection{From Masses to Wilson Coefficients}  \label{sec:MassResOpBund}

To go from a finite spectrum of masses to Wilson coefficients of our EFT local operators at four-points, we simply expand in the  large effective cutoff limit.  If the spectrum of masses is given in terms of $\mu_k$ with mass scale $\Lambda$, 
\begin{equation}
M=\prod_{k=1} \frac{P_k(\sigma_2, \sigma_3)}{(s/\Lambda-\mu_k)(t/\Lambda-\mu_k)(u/\Lambda-\mu_k)} 
\end{equation}
then we simply series expand $\Lambda\to\infty$ and read off the Wilson coefficients as functions of scaleless $\mu_k$ order by order in some $\alpha'=1/\Lambda$. Of course, generic multiplication by terms of order:
\begin{equation}
M=\prod_{k=1} \frac{f_k}{(s/\Lambda-\mu_k)(t/\Lambda-\mu_k)(u/\Lambda-\mu_k)} 
\end{equation}
where $f_k$ is independent of kinematics, certainly establishes a tower of massive modes, but are subject to whatever massless factorization properties are present in the IR theory. One might worry about contact terms of the form $\sigma_2^m/s$, which suggest the presence of higher-spin massless exchange in the EFT. In this paper, we postpone judgement on the matter for the following reasons:
\begin{itemize}
 \item Such dubious channels can be disposed of by the other theory appearing in the double-copy with, e.g.~the nonlinear sigma model (NLSM) is famous for absorbing channels in the double-copy. We will use NLSM as an exemplar of color-dual model building with massive resonance in the next section.
 \item If $\sigma_2^m/s$ issues are resolved via double-copy with all theories of interest, this can be suggestive of applying adjoint-type double-copy construction where symmetric $d^{abc}$ type double-copy results in identical well-behaved double-copy theories but with more physical looking ingredients \cite{Carrasco:2022jxn,Carrasco:2023qgz}.  This strikes us as similar to using a gauge with spurious poles that cancel in the end --- perhaps aesthetically unappealing, but not actually problematic unless one has real issues in the final product.
 \end{itemize}
 That said, $\sigma_2$, glitches can be entirely disposed of if one is willing to start with an IR theory that is proportional to $\sigma_3$. Another choice of polynomial residue sufficient for canceling unwanted higher spin modes at four-point would be, $P_n(\sigma_2,\sigma_3)= \mu_n^3-\mu_n \sigma_2+\sigma_3$. Given such a choice of numerator, the massive spectra are related to the Wilson coefficients of the exponential factors at four-point as follows:
\begin{eBox}
\be  \label{ExpCoefficientsToMasses}
c_{(1,m)} = \sum_{n=1}^\infty \frac{1}{\mu^{2m+3}_n}
\ee
\end{eBox}
Clearly, when $\mu_n = n$, the Wilson coefficients that we have shown exponentiate at four-point, just reproduce the zeta values of the single-value promotion, $c_{(1,m)} = \zeta_{2m+3}$. 

We should note that simply engaging with a finite spectrum of massive modes, while it may lead to incredibly well behaved UV scaling, could introduce problematic behavior like violating positivity bounds~\cite{Adams:2006sv}\footnote{See for example refs.~\cite{deRham:2017avq,Bellazzini:2019xts,Bellazzini:2020cot,Bern:2021ppb,Creminelli:2022onn} for recent applications of imposing positivity constraints.}, or disturbing partial wave unitarity upon resummation\footnote{A venerable consistency constraint~\cite{Martin:1962rt,Goddard:1972iy,Soldate:1986mk,Aoki:1990yn} but see also e.g. refs.~\cite{Caron-Huot:2016icg,Green:2019tpt,Arkani-Hamed:2022gsa} for recent discussions.}. Of course, one is free to keep adding and tuning modes ad-infinitum until one lands on an EFT that looks like a healthy UV theory, or be content with an EFT that allows one to describe the relevant physics at hand to desired accuracy. Either way, empirically there is a much more pressing problem.

The mass parameters are \textit{non-perturbative}, requiring information of the full theory at energy scales beyond the UV cutoff scale, $\Lambda=1/\alpha'$.  However, often we only have empirical access to the Wilson coefficients, $c_k$, that are valid below the cutoff scale. Solving for the mass spectrum analytically runs into an inverse problem where we must determine an infinite number of degrees of freedom, $\mu_n$, from finite number of outputs, $c_{(1,m)}$ (up to some fixed order in $\alpha'$), and thus may seem forever out of reach from an amplitudes-based perturbative bootstrap. Fortunately, there are incredibly well-developed mathematical tools that we can use to extract non-perturbative information about the spectrum, directly from the Wilson coefficients of color-dual operators appearing in \eqn{allMultiplicityFormula}. We turn to these methods now.

\subsection{From Wilson Coefficients to Masses}\label{padeSection}
In the remainder of this section we will introduce Pad\'{e} approximants as an analytic tool for probing the \textit{non-perturbative} spectra of double-copy consistent amplitudes with generic Wilson coefficients. 
A Pad\'{e} approximant is a rational function of the form,
\be\label{Pade}
\mathcal{R}_{[m,n]}(x) \equiv \frac{A_m(x)}{B_n(x)}\,,
\ee
where $[m,n]$ is the degree of the polynomials, $A_m$ and $B_n$, respectively:
\be\label{polys}
A_m(x) = \sum_{j=0}^m a_j x^j, \quad B_n(x) = 1+\sum_{j=0}^n b_j x^j\,.
\ee
The coefficients of $\mathcal{R}_{[m,n]}$ are defined in order to match the first $m+n$ derivatives when expanded around a point, $x=x_0$,
\be\label{PadeDef}
f^{(k)}(x_0) = \mathcal{R}^{(k)}_{[m,n]}(x_0) \qquad k\leq m+n\,.
\ee
Pad\'{e} approximants have been used widely in the resurgence literature for extracting information about non-perturbative physics, like instantons, when only pertubative information is available. For background, we direct the reader to refs. \cite{Costin:2020hwg,Costin:2020pcj,Dunne:2021acr,Dunne:2022esi}. For our purposes, we will use them to study the behavior of mass spectra in generic classes of consistent mass resonance double-copy models, even when the closed form resummation of the amplitude is not directly computable. 

As straightforward example that demonstrates the power of these ideas in the context of effective field theory, let us consider a scalar effective field theory that looks like a chiral perturbation theory \cite{Weinberg:1978kz}, with two resonant masses beyond the $\alpha'=1$ scale of the pion decay width. To demonstrate the reach of these non-perturbative methods, we introduce an order of magnitude separation between resonances with $\mu_1=1$ and $\mu_2=10$. In our framework the purely IR amplitude, $\mathcal{A}_\text{IR}$, is just that of the nonlinear sigma model (NLSM), a known color-dual theory~\cite{Chen2013fya,Cachazo:2014xea,Cheung2016prv,Cheung:2017yef}, which is the leading order contribution in chiral Lagrangian \cite{Weinberg:1978kz}. The four-point amplitude is simply:
\begin{equation}
\label{aIRNLSM}
\mathcal{A}_\text{IR} = f^{a_1a_2b}f^{b a_3 a_4} (s_{23}-s_{13}) +
f^{a_4a_1b}f^{b a_2 a_3} (s_{12}-s_{13})+
f^{a_3a_1b}f^{b a_4 a_2} (s_{23}-s_{12})\,.
\end{equation}
To recover the pions of the Standard Model, one simply treats the structure constants as describing some flavor symmetry, speficying $SU(2)$ for the desired isospin symmetry of low energy QCD. We will embed the UV mass spectrum encoded by,
\begin{equation}
M^{\text{2-mode}}=\frac{1\times10^3}{(s\alpha'-1)(t\alpha'-1)(u\alpha'-1)(s\alpha'-10)(t\alpha'-10)(u\alpha'-10)}\,.
\end{equation}
Again, $\alpha'=1/\Lambda$, where $\Lambda$ is the cutoff scale and $\alpha'$ is the pion decay width for the NLSM. For the purposes of this exercise, we take ourselves as living deep in the IR and so only have experimental access to the Wilson coefficients measured in some fixed-angle scattering,  taking $u=-s-t$ and $t=-s/2$, up  to some finite precision.

Let us consider the difference between what we can infer with Pad\'{e} approximants if we have access to 6 vs 10 Wilson coefficients, to a precision of four significant digits\footnote{Perhaps we should emphasize that this does not come for free.  In order to measure the Wilson coefficients to such precision at $\mathcal{O}(\alpha')^{10}$ deep in the IR requires astoundingly accurate experiments.}.  The relevant expansions are subsets of the series expansion, 
\begin{align}
\label{nlsmExpansion}
\mathcal{A}&=\mathcal{A}^{\text{IR}}\times
 \left[ 1.000+1.010 (\alpha ')^2 \sigma_2+1.001 (\alpha ')^3 \sigma_3
 +1.010 (\alpha ')^4 \sigma_2^2+2.01 (\alpha ')^5 \sigma _2 \sigma _3 \right. \nn \\
&~+(\alpha ')^6 \left(1.010 \sigma _2^3+1.001 \sigma _3^2\right)+3.02 (\alpha')^7 \sigma _2^2 \sigma _3+(\alpha ')^8 \left(1.010 \sigma _2^4+3.01 \sigma _2 \sigma
_3^2\right)\nn\\
&~\left.+(\alpha ')^9 \left(4.03 \sigma _2^3 \sigma_3+1.001 \sigma_3^3\right) 
+(\alpha ')^{10} \left(1.01 \sigma _2^5+6.03 \sigma _2^2 \sigma _3^2\right)+
\mathcal{O}(\alpha')^{11} \right] \,.
\end{align}

\begin{figure}[t]
\includegraphics[width=0.47\textwidth]{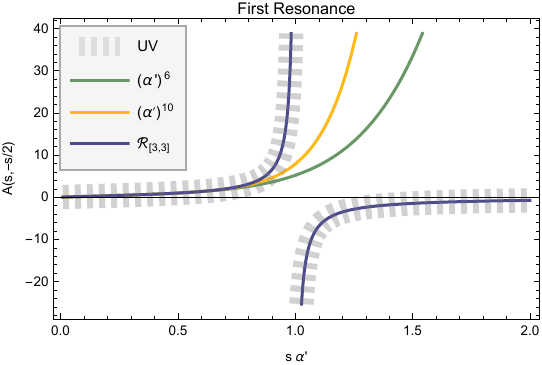} \,
\includegraphics[width=0.47\textwidth]{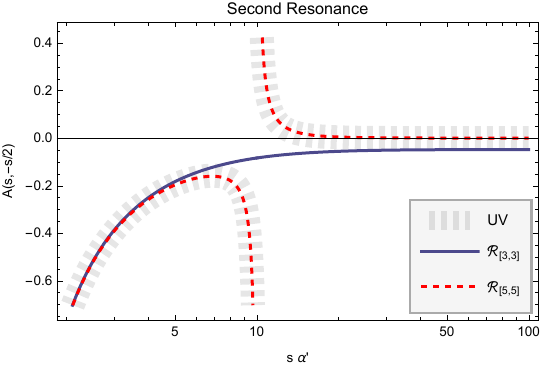}
 \caption{Expanded series expansions, $\mathcal{O}(\alpha')^6$ and $\mathcal{O}(\alpha')^{10}$, as well as Pad\'{e} approximant $\mathcal{R}_{[3,3]}$ to the $\mathcal{O}(\alpha')^{6}$ expansion (left) plotted against the fully resummed UV function.   No matter how high the perturbative expansion, it will never capture even the first emergent resonance, but the behavior is spectacularly captured in the $\mathcal{R}_{[3,3]}$ approximant. On the second panel (right), we show the behavior of Pad\'{e} approximants  $\mathcal{R}_{[3,3]}$ to the $\mathcal{O}(\alpha')^6$ series expansion and $\mathcal{R}_{[5,5]}$ to the $\mathcal{O}(\alpha')^{10}$ series expansion near the domain of the second resonance. We find that a relatively small number of Wilson coefficients allow us in $\mathcal{R}_{[5,5]}$ to recover the full  spectrum of massive modes.}\label{twoResonanceModel}
\end{figure}

Of course any perturbative expansion will fail to catch even the first resonance.  Even with access to only four-significant digits, the $\mathcal{R}_{[3,3]}$ approximant to the $\mathcal{O}(\alpha')^6$ expansion  captures the first resonance, and the  $\mathcal{R}_{[5,5]}$  approximant to the $\mathcal{O}(\alpha')^{10}$ expansion manages to encode  both resonances.  We plot the fixed-angle predictions for the $A(1234)$ ordered amplitude\footnote{The coefficient of $f^{a_1a_2b}f^{b a_3 a_4}$ after using the Jacobi identity to rewrite $f^{a_4a_1b}f^{b a_2 a_3}=f^{a_1a_2b}f^{b a_3 a_4}-f^{a_3a_1b}f^{b a_4 a_2}$ in \eqn{nlsmExpansion} as per \sect{background}.} in \fig{twoResonanceModel}.

 As we discuss in the conclusion, leveraging non-perturbative data from such a small number of Wilson-coefficients has a number of potential opportunities both for empirical expectations for new signals as well as formal insight relating non-perturbative behavior of theories participating in the double-copy web of theories.

\newpage\section{Conclusions}\label{conclusions}
In this work we have demonstrated two mechanisms for understanding UV behavior of scattering amplitudes directly from perturbative information in the IR -- double-copy consistency of effective field theory, and Pad\'{e} approximants for fixed-angle 2-to-2 scattering amplitudes. To properly frame these approaches, we provide below a brief outline of the main results of the paper and future directions made possible by this work.

\subsection{Summary } \label{summary}
\paragraph{Menu of Massive Resonance} We began with an overview of the interplay between massive resonance and double-copy constructible theories studied in the literature from $\text{YM}+\text{F}^3$ theory in \sect{bosonicHint}, to string amplitudes and $Z$-theory in \sect{StringyExamples}. We provided an argument for why one might expect color-kinematics duality to impose all-order constraints on effective operators that in principle could resum to non-local kinetic operators in the UV. In \fig{colorDualPyramid}, we introduced the color-dual factorization pyramid, where gauge field operators at fixed orders in $\alpha'^{\,n}$,
\be
\mathcal{O}(\alpha'^n):\quad A^{2n+4}\leftrightarrow \partial^2 A^{2n+2}\leftrightarrow \partial^4 A^{2n}\leftrightarrow\cdots\leftrightarrow  \partial^{2n+2} A^{2}\,,
\ee
are mixed via color-dual constraints on kinematic numerators (or partial amplitudes). Then, factorization bundles together remaining operators higher up on the pyramid at all orders $\alpha'^{m<n}$. Through this bundling, one might expect massive resonances to appear due to modified kinetic terms, $\mathcal{L}_{\text{kin}}\supset \prod_{k} (\partial^2-\mu_k^2)A^2$. We thus proposed a simple paradigm for encoding massive resonance through all order towers of higher derivatives that live in massless color-dual numerators of the form,
\begin{eBox}
\be
\mathcal{M}^{\text{mass}} = \sum_{g}\frac{n_g N^{\text{HD}}_g}{d_g}\,.
\ee
\end{eBox}
Motivated by the higher derivative bootstrap of $\text{YM}+\text{F}^3$ studied \cite{Carrasco:2022lbm}, we then pursued a bootstrap to determine to what extent the Wilson coefficients of higher-derivative operators are constrained by double-copy consistency. 
\paragraph{Color-Dual Bootstrap}
In \sect{sec:SVBootstrap}, we began by studying the promotion of a color-dual four-point contact, $\sigma_3^n A_4$, with Wilson coefficients, $c_{(n,0)}$, to five-point amplitudes, $A_5^{(n)}$, under double-copy consistency. We found that through third order, that the amplitude is completely determined, and exhibits an iterative structure in terms of a higher-derivative transformation matrix, $A_5^{(n)}=\Omega^n A_5$. Requiring that the six-point amplitude is color-dual and factorizes to $\Omega^n A_5$, imposes a constraint on the split factorization channel of \eqn{6ptMatrixSplitCutResultofCKAllorder}. Double-copy consistency thus requires that the four-point contact exponentiates:
\be
c_{(n,0)} = \frac{1}{n!}c_{(1,0)}^n
\ee
Guided by the structure of this bootstrap, in \eqn{genColorDualContact} through \eqn{genCutConstraint}, we provided an all-order procedure for generalizing this exponential behavior to all-multiplicity by imposing a series of $\text{N}^k\text{MCut}$ conditions -- finding that IR color-dual amplitudes must in general exponentiate any higher-derivative contact operator consistent with the duality:
\begin{eBox}
\be \label{allMultiplicityFormula2}
A^{\text{UV}}_n = \prod_{k}\exp\left[c_k\Omega^{\text{DCC}}_k \right]A^{\text{IR}}_n\,,
\ee
\end{eBox}
where $\Omega^{\text{DCC}}_k$ are double-copy consistent higher derivative matrices that map color-dual amplitudes to color-dual contacts at higher orders in the effective expansion. In doing so, we have demonstrated that double-copy consistency alone could be sufficient to introduce kinematic factors that soften the UV behavior of gravity, among all other theories that are double-copy constructible.  

\paragraph{UV Spectra via Pad\'{e}} 
Finally, to resolve the inverse problem posed by the expression of \eqn{ExpCoefficientsToMasses}, we invoked Pad\'{e} approximants used widely in the resurgence literature \cite{Costin:2020pcj,Costin:2020hwg,Dunne:2021acr,Dunne:2022esi},
\begin{eBox}
\be
R^{(k)}_{[m,n]}(x_0)=f^{(k)}(x_0) \qquad k\leq m+n\,,
\ee
\end{eBox}
where $f^{(k)}$ is the $k$-th derivative of $f(x)$. This analytic continuation can be used as a mechanism for recovering non-perturbative information about the mass spectrum, directly from the Wilson coefficients of \eqn{allMultiplicityFormula}. In doing so we show how increasing orders in Wilson coefficients allow us to gain further reach in the spectrum of the UV theory as per \fig{twoResonanceModel}, directly from Wilson coefficients in the IR.  

 Before discussing some specific future directions, we will comment about the potential implications of the color-dual factorization pyramid for novel representations of color-dual loop-level integrands.

\subsection{Pyramid Implications for Loop Corrections}  \label{loops}
 If we take seriously the possibility that a $D$-dimensional classification of the kinematic algebra requires one to introduce to non-local field theory operators, then restricting ourselves to just consider $F^2$ operators at Yang-Mills mass-dimension might prove too restrictive for constructing all-loop color-dual integrands. Here we sketch an argument for why one might expect higher dimensional operators to force themselves into the conversation starting at two-loop. 

Suppose we have a color-dual three-point operator that manifests Jacobi-like relations at all multiplicity in arbitrary dimensions, $\mathbb{V}_{(abc)}$. As we argued above, consistency conditions between higher derivative contact diagrams could introduce non-local propagators with modified kinetic terms. These would appear as corrections to the propagators -- yielding the following sum over physical states:
\be  \label{vertexDef}
\mathbb{V}_{(ab|k)} \cdot \mathbb{P}_{(k)}\cdot \mathbb{V}_{(-k|cd)} 
\ee
where in general, our state-projector operator could be some string-like vertex operator with a tower of massive-resonant modes, like those appearing in the amplitudes of $Z$-theory:
\be  \label{HigherSpinSum}
\mathbb{P}_{(k)} \sim\frac{\eta^{\mu\nu}}{k^2} + \sum_{n=0}^\infty \sum_{j=1}^n \frac{P^{\mu_1\mu_2...\mu_j}_{\nu_1\nu_2...\nu_j}}{(k^2-\mu_n)}
\ee
However, as we have argued in the text, color-dual theories with massive resonances should be captured as higher derivative corrections in the cubic graphs. After absorbing these massive residues into the definition of the vertices of kinematic numerators, we could expect to see interactions resembling,
\be \label{massAbsorbedVertex}
\mathbb{V}_{(123)}\rightarrow \mathbb{V}^{\text{YM}}_{(123)}+\sum_{n_1,n_2,n_3} \frac{B_1^{\otimes({n_{1}})}\cdot B_2^{\otimes({n_{2}})}\cdot B_3^{\otimes({n_{3}})}}{(k_1^2 -\mu_{n_1})(k_2^2 -\mu_{n_2})(k_3^2 -\mu_{n_3})}
\ee
where $B^{\otimes(n)}$ is an on-shell state tensor valued operator with upto spin-$n$ modes. When inserting only on-shell vector states of Yang-Mills, in the $\alpha'\rightarrow 0$ limit, we know that this must recover the normal on-shell Yang-Mills vertex:
\be  \label{fieldTheoryVertexLimit}
 \lim_{\alpha'\to 0}\mathbb{V}_{(123)}\cdot \varepsilon_i \rightarrow \mathbb{V}^{\text{YM}}_{(123)}\cdot \varepsilon_i 
\ee
At both tree-level and one-loop, the basis diagrams for color-dual theory expose every internal vertex to at least one on-shell state:
\be \label{NgonHalffLadder}
\halfLadder{2}{3} \qquad \qquad \nGon
\ee
However, at two-loop, the off-shell vertices become relevant in the basis diagrams of a color-dual gauge theory:
\be \label{TwoLoopBasis}
\doubleBox \qquad \qquad \pentaTri
\ee
The red vertices are completely off-shell, and thus there would be no external state to temper any non-local denominator introduced by $D$-dimensional double copy consistency. Since purely non-vector modes would be allowed to contribute to the state-sum of the two-loop basis graph, color-dual numerators at two-loop and beyond could be constructed from $\textit{non-local}$ kinematic numerators of rational functions, rather than of polynomial kinematics that one would expect from local Feynman rules.  Off-shell local color-dual numerators of minimal powercounting was excluded for Yang-Millls at two-loops and four-points in arbitrary dimensions in ref.~\cite{Bern:2015ooa}. This exclusion was recently confirmed and extended by relaxing all power-counting assumptions by recent work in collaboration with one of the authors~\cite{Edison:2023ulf}. A by now venerable example of non-local representations at two-loop can be found in \cite{Mogull:2015adi}, which studied five-point Yang-Mills integrands for a specific choice of 4D external states. 
\subsection{Future Directions}  \label{futureDirections}

We see many future directions and applications of this work. Possibly the most exciting is the prospect for phenomenological model building. The results of our bootstrap suggest that the space of double-copy consistent amplitudes is necessarily softened at high energy, while providing freedom to tune Wilson coefficients beyond the zeta values of string theory. In \sect{sec:MassResOpBund}, we shared our  framework for using Pad\'{e} approximant to study massive spectra above the cutoff scale. Better understanding the sensitivity of UV physics on the IR operators would allow the construction of bespoke UV completions. It would be worth understanding how our double-copy consistent operators, $\Omega_k^{\text{DCC}}$, when appropriately tuned match the spectral constructions of \cite{Cheung:2023uwn}. 

To this end, a natural application of our results towards model building would be in the low energy confining regime of QCD.  One could use the massive resonance framework of \eqn{massiveRes} to construct pionic scattering models that incorporate excited meson resonances, and then promote the spectrum to higher-multiplicity through exponentiated contacts of \eqn{allMultiplicityFormula}. Indeed, a large class models in the web of color-dual theories~\cite{BCJreview}, not just low energy QCD, could be uplifted to consistent higher-multiplicity massive resonance scattering in this way.

Beyond the possibility of model building, we should emphasize that the exponential structure we have identified only takes into account constraints placed by color-kinematics and \textit{split factorization}. However, these constraints do not preclude the possibility of additional relations between Wilson coefficients from more factorization channels at higher multiplicity. Consider for example an eight-point factorization channel that mixes four-point contacts at different orders:
\be
{\lim_{{s_{123} \rightarrow 0 \atop s_{678} \rightarrow 0}}}\,\,\factGraphsEight{\large$A_8^{(13)}$}\!\!=\frac{1}{s_{123} s_{678}}\left[
\begin{split}
&\factGraphsTrip{nhpBlue4}{nhpBlue2}{$\sigma_3$}{$\sigma_5$}{$\sigma_5$}{nhpBlue2}
\\
&\factGraphsTrip{nhpBlue1}{nhpBlue4}{$\sigma_7$}{$\sigma_3$}{$\sigma_3$}{nhpBlue4}\end{split}+\begin{split}
&\factGraphsTrip{nhpBlue2}{nhpBlue4}{$\sigma_3$}{$\sigma_5$}{$\sigma_5$}{nhpBlue2}
\\
&\factGraphsTrip{nhpBlue4}{nhpBlue1}{$\sigma_3$}{$\sigma_7$}{$\sigma_3$}{nhpBlue4}\end{split}+\cdots
\right]
\ee
where $\sigma_k = c_{(k)}A^{(0)}_4 (s^k+t^k+u^k)$. By considering additional all-order constraints, our exponentiated contacts could further conspire at overlapping orders in the effective $\alpha'$ expansion. Better understanding whether there are additional relations between the Wilson coefficients of \eqn{allMultiplicityFormula} we see as an important direction of future study, that could give further understanding about the uniqueness of stringy dynamics. 

Absent additional constraints, our bootstrap at the very least offers an IR perspective on why certain terms are excluded from the low energy effective expansions in the closed string. For example, consider the multiple zeta value, $\zeta_{3,5}$, that appears in the open string expansion, but is absent in the closed string \cite{Broedel:2013tta}. From a UV perspective, $\zeta_{3,5}$ is not single-valued and thus cannot be constructed from the spherical integration of closed string amplitudes.  Fascinatingly, we can now identify an IR reason from our bootstrap for why this term must be excluded from the closed string expansion. 

Consider, by way of example, the $\mathcal{O}(\alpha'^8)$ sector at five-point, the order at which one would expect  $\zeta_{3,5}$ to appear.  At this order in the open string expansion, $\zeta_{3,5}$ weights a contact of the form, $\Omega^{\text{OS}}_{5,8}\sim \zeta_{3,5} \left(\Omega^{(1,0)}\Omega^{(1,1)}-\Omega^{(1,1)}\Omega^{(1,0)}\right)$, where $\Omega^{(m,n)}$ are the five-point higher-derivative matrices that we defined in \eqn{NptDCCAmpAnsatz}, and $\Omega^{\text{OS}}_{5,8}$ just indicates the open string five-point contact that appears at $\mathcal{O}(\alpha'^8)$. The only permissible factorization channel at this order is the following:
\be
{\lim_{s_{45}\to0} }\,\,\factGraphsFive{\large$A_5^{(8)}$}{1}{2}{3}{4}{5}\!\!=c_{(2,1)}\frac{\sigma_{3}^2 \sigma_2}{s_{45}}\,\factGraphsFFF{1}{2}{3}{4}{5}{\large$ A_4$}{nhpBlue4}{\large$A_3$}
\ee
After imposing color-kinematic constraints at five-point, and demand consistent factorization down to four-point, we find that there are only two additional color-dual five-point contacts. Let us  call these contacts $\Omega_{5,8}^{1}$ and $\Omega_{5,8}^{2}$, where $\Omega_{m,n}$ is a $m$-point contact at $\mathcal{O}(\alpha'^n)$. Intriguingly, one finds that $\Omega^{\text{OS}}$ is not compatible with any linear combination of $\Omega_{5,8}^{1}$ and $\Omega_{5,8}^{2}$. Thus we find that the $\zeta_{3,5}$ Wilson coefficient is restricted from the closed string expansion because its operator  is excluded by double-copy consistency.  An obviously compelling future direction is to build a detailed understanding of which IR operators in the closed and open string expansions are prohibited by double-copy consistency, thereby understanding  entirely IR  constraints on effective operators .

Finally, it would be worth further studying the extent to which resurgence can be leveraged to make UV predictions from IR physics. One could start by understanding the relationship between partial wave unitarity \cite{Caron-Huot:2016icg,Arkani-Hamed:2022gsa} applied to massive residues of Pad\'{e} approximants, and positivity bounds on Wilson coefficients from the optical theorem \cite{Adams:2006sv}. Indeed, utilizing analytic continuations of effective field theory beyond the perturbative sector could be fruitful in further bounding the UV directly from IR data. Furthermore, it is clear that leveraging resurgence methods in an EFT context should serve as a mechanism for identifying potential energy scales for new physics directly from a finite number of Wilson coefficients. It does not escape us that applying resurgence methods to Wilson coefficients gleaned from loop-level effective predictions, directly constructed via  on-shell methods, could represent a rather novel amplitudes path towards probing non-perturbative physics in theories both of phenomenological and formal interest. This has the potential to expose shared non-perturbative structure within the double-copy web of theories.

\begin{acknowledgments} The authors would like to thank Rafael Aoude, Sasank Chava, Lance Dixon, Andr\'{e} de Gouv\^{e}a, Gerald Dunne, Alex Edison, Kays Haddad, Karol Kampf, Ian Low, James Mangan, Frank Petriello, Paolo Pichini, Radu Roiban, Oliver Schlotterer,  Aslan Seifi, Jaroslav Trnka, and Suna Zekio\u{g}lu for insightful conversations, related collaboration, and encouragement throughout the completion of this work.  We would especially like to thank Gerald Dunne, Oliver Schlotterer, and Radu Roiban for very valuable feedback on earlier versions of the manuscript. The completion of this manuscript benefited from the hospitality of NORDITA during the program ``Amplifying Gravity at All Scales", and ICTP-SAIFR during the program ``Gravitational Waves meet Amplitudes in the Southern Hemisphere.'' This work was supported by the DOE under contract DE-SC0015910 and by the Alfred P. Sloan Foundation. N.H.P. acknowledges the DOE Office of Science Graduate Student Research (SCGSR) program under contract DE‐SC0014664, the Northwestern University Amplitudes and Insight group, the Department of Physics and Astronomy, and Weinberg College for their generous support. 
\end{acknowledgments}

\appendix

\newpage\section{Hypergeometric Residues}  \label{HypergeometricRes}
Below we provide some identities that we use to recover the residues of the hypergeometric functions appearing in five-point $Z$-theory amplitudes. The ${}_3F_2$ hypergeometric function can be expressed as an infinite sum over Pochhammer symbols,

\be  \label{3F2Def}
{}_3F_2\left[
{a_1,\,a_2,\,a_3\atop
a_4,\,a_5}
;z\right]= \sum_{n=0}^\infty \frac{(a_1)_n(a_2)_n(a_3)_n}{(a_4)_n(a_5)_n} \frac{z^n}{n!} \,,
\ee
where $(x)_n = \Gamma(x+n)/\Gamma(x)$. In the text we found that the fully resummed $Z$-theory $(n-3)!$ basis amplitudes can be expressed in terms of a single functional with different kinematic arguments, depending on whether the color ordering is planar or non-planar with respect to the Chan-Paton factors:
\begin{align} \label{Z5Structure}
\Zpent{5}{1}{2}{3}{4}&= Z_5(-s_{34},-s_{45},-s_{15},-s_{12},-s_{23})\,,
\\
\Zpuff{5}{1}{2}{3}{4}& =-  Z_5(1-s_{34},-s_{45},-s_{15},1-s_{12},-s_{23})\,.
\end{align}
Here we have defined the five-point $Z_5$-function using a particular ordering of ${}_3F_2$ as follows,
\be  \label{Z5Def}
Z_5(a_1,a_2,a_3,a_4,a_5) \equiv {}_3F_2\left[{
\,a_{1},\,a_{4},\,a_{23}-a_5\atop a_{12},\,a_{34}};1\right] \frac{\Gamma(a_1)\Gamma(a_2)\Gamma(a_3)\Gamma(a_4)}{\Gamma(a_{12})\Gamma(a_{34})}\,,
\ee
and employed the shorthand notation $a_{i_1i_2...i_n} = a_{i_1}+a_{i_2}+\cdots +a_{i_n}$. Since ${}_3F_2 \rightarrow 1$ when anyone of the top arguments vanishes, it is clear that $Z_5$ factorizes to the appropriate four-point Veneziano amplitudes in the $a_1,a_4\rightarrow 0$ limits. However, as noted in the text, the other factorization channels are obscured when defined in this way. To elucidate the remaining pole structure, one can apply a Thomae transformation to the $Z_5$ function,
\be  \label{Thomae}
\frac{(x_{123})!}{(x_{13})!(x_{23})!} {}_3F_2\left[
{1+x_{123},\,-x_4,\,-x_5\atop
1+x_{13},\,1+x_{23}}
;1\right]
=
\frac{(x_{345})!}{(x_{34})!(x_{35})!}  {}_3F_2\left[
{1+x_{345},\,-x_1,\,-x_2\atop
1+x_{34},\,1+x_{35}}
;1\right]\,.
\ee
By making the following variable replacements, one can show that the $Z_5$ is cyclically invariant in its arguments:
\begin{align} \label{HyperGRotate}
1+x_{123} = a_1 \quad &\Rightarrow \quad Z_5(a_1,a_2,a_3,a_4,a_5) = Z_5(a_2,a_3,a_4,a_5,a_1)
\\
1+x_{123} = a_4 \quad &\Rightarrow \quad Z_5(a_1,a_2,a_3,a_4,a_5) = Z_5(a_5,a_1,a_2,a_3,a_4)
\end{align}
As such, we can use this relationship between different cyclic orderings of $Z_5$ do derive the $a_i\rightarrow 0$ limits of the hypergeometric function appearing in Z-theory amplitudes:
\begin{align}  \label{HyperGLimits}
\lim_{a_{1}\rightarrow 0} {}_3F_2\left[{
\,a_{1},\,a_{4},\,a_{23}-a_5\atop a_{12},\,a_{34}};1\right] &= 1
\\
\lim_{a_2\rightarrow 0} {}_3F_2\left[{
\,a_{1},\,a_{4},\,a_{23}-a_5\atop a_{12},\,a_{34}};1\right] &= {}_2F_1\left[{
a_{4},\,a_{3}-a_5\atop a_{34}};1\right]  = \frac{\Gamma(a_5) \Gamma(a_{34})}{\Gamma(a_3) \Gamma(a_{45})}
\\
\lim_{a_3\rightarrow 0} {}_3F_2\left[{
\,a_{1},\,a_{4},\,a_{23}-a_5\atop a_{12},\,a_{34}};1\right] &={}_2F_1\left[{
a_{1},\,a_{2}-a_5\atop a_{12}};1\right]  = \frac{\Gamma(a_5) \Gamma(a_{12})}{\Gamma(a_2) \Gamma(a_{15})}
\\
\lim_{a_{4}\rightarrow 0} {}_3F_2\left[{
\,a_{1},\,a_{4},\,a_{23}-a_5\atop a_{12},\,a_{34}};1\right] &= 1
\\
\lim_{a_5\rightarrow 0} {}_3F_2\left[{
\,a_{1},\,a_{4},\,a_{23}-a_5\atop a_{12},\,a_{34}};1\right] &= \frac{1}{a_5} \frac{\Gamma(a_{12})\Gamma(a_{34})}{\Gamma(a_{1})\Gamma(a_{23}) \Gamma(a_{4})}+\mathcal{O}(1)
\end{align}
Together these limits of the ${}_3F_2$ hypergeometric function give the desired factorization properties endowed in the fully resummed $Z$-theory amplitudes studied in the text. While in the text we have just studied the $s_{12}$ and $s_{34}$ factorization channels at five-point -- the remaining residues can simply be recovered through functional relabelings due to the cyclicity of $Z_5$.

\bibliography{Refs_CMRDC.bib}

\end{document}